\documentclass{article}
\usepackage[dvips]{graphicx}
\usepackage{amsfonts}
\usepackage{amsmath}
\title{The $\mathbb{R}P^{2}$ Valued Sigma and Baby Skyrme Models}
\author{Matthew D E Szyndel}

\begin{document}

\begin{center}
\vspace*{2cm}
{\Large \bf The $\mathbb{R}P^{2}$ Valued Sigma and Baby Skyrme Models}

\vspace*{1cm}
Matthew D E Szyndel\footnote{email: m.d.e.szyndel@durham.ac.uk}

\vspace*{0.5cm}
\begin{footnotesize}
Centre for Particle Theory,\\ Mathematical Sciences,\\ University of Durham,\\ Durham,\\ DH1 3LE,\\ UK.
\end{footnotesize}
\end{center}

\begin{abstract}
We investigate the sigma and baby Skyrme models with an $\mathbb{R}P^{2}$
target space. We compare these models to models with an $S^{2}$ target
space. We investigate the interactions between solitons and defects in the
$\mathbb{R}P^{2}$ sigma model.
\end{abstract}

\section{Introduction}

The Skyrme model has long been of interest as an effective field theory of
nucleons \cite{skyrorig}. The 2 dimensional baby Skyrme model allows us to study a more
tractable analogue of the Skyrme model \cite{review}. The related O(3) sigma model is
applicable to some condensed matter systems. Nematic liquid crystals are a
condensed matter system where the target space is not $S^{2}$ but
$\mathbb{R}P^{2}$. In this paper we describe our work on the variants of the
baby Skyrme model and O(3) sigma model using $\mathbb{R}P^{2}$ as a target
space.

\subsection{Models}

The O(3) sigma model describes maps from $\mathbb{R}^{2}$ (the
``physical'' space) to $S^{2}$ (the ``target'' space). $\mathbb{R}^{2}$
is the normal Euclidean plane, whereas $S^{2}$ is the 2 dimensional surface
of a unit sphere embedded in 3 Euclidean dimensions. The lagragian of the
O(3) sigma model is given by
\begin{equation} \label{siglag}
\mathcal{L}= \frac{1}{4}\partial^{\mu}\vec\phi\cdot\partial_{\mu}\vec\phi,
\end{equation}
where $\vec\phi \in \mathbb{R}^{3}$ such that $\vec\phi \cdot \vec\phi = 1$.
The condition that $\vec\phi \mapsto \vec\phi_{vac}$ as $r \mapsto \infty$,
where $r$ is spatial radius, is usually imposed to keep the action finite.
This effectively compactifies the physical space from $\mathbb{R}^{2}$ to
$S^{2}$. This in turn implies that the configurations $\vec\phi$ must fall
into disjoint classes characterised by the elements of the second homotopy
group $\pi_{2}(S^{2}) \cong \mathbb{Z}$. Field configurations which fall into
any of the classes corresponding to non-trivial elements of $\pi_{2}(S^{2})$
describe `lumps' of energy. These lumps are not stable, and as a result are not
true solitons. This instability is due to the conformal invariance of the
model -- the lumps have no intrinsic scale, and so may change scale without
any energetic penalty. This scale invariance may be removed by the addition
of new terms to the Lagrangian, as seen below.

The first model we examined was the baby Skyrme model with the so called
`new' potential term. The Lagrangian is
\begin{equation} \label{skylag}
\mathcal{L}= \frac{1}{4}\partial^{\mu}\vec\phi\cdot\partial_{\mu}\vec\phi
 + \theta_{1}\frac{1}{4}((\partial^{\mu}\vec\phi\cdot\partial^{\nu}\vec\phi)
(\partial_{\mu}\vec\phi\cdot\partial_{\nu}\vec\phi)
-(\partial^{\mu}\vec\phi\cdot\partial_{\mu}\vec\phi)^{2})
+\theta_{2}\mathcal{L}_{V},
\end{equation}
where $\theta_{1}$ and $\theta_{2}$ are real positive constants. The term
containing $\theta_{1}$ is the so called (2 dimensional) Skyrme term. This term
tends to cause solitons to broaden, and is the only possible Lorentz covariant
term of fourth order with only first order time derivatives. The term containing
$\theta_{2}$ is normally refered to as the potential term. This term is chosen
so that it acts to shrink the size of the soliton. Consequently a stable equilibrium
may be reached where the effects of the Skyrme term and potential term balance. A
number of different potential terms have been studied \cite{review,tom}, but our
$\mathbb{R}P^{2}$ target space forces us to use a potential term which is 
invariant under the transformation $\vec \phi \mapsto -\vec \phi$. A good
choice would be the so called `new' potential
\begin{equation} \label{newpot}
\mathcal{L}_{V}=\frac{1}{4} (1-(\vec\phi\cdot\vec\phi_{vac})^{2}).
\end{equation}

\subsection{The Geometry and Topology of $\mathbb{R}P^{2}$}
\label{geomtop}

\subsubsection{Geometry}
\label{geom}

$\mathbb{R}P^{2}$ may be thought of as a hemisphere, and so may be
parameterised by a unit 3-vector $\vec\phi$ where $\vec\phi(x) \equiv -\vec\phi(x)$.
$\mathbb{R}P^{2}$ may also be described locally by a 2-vector $\vec\chi$
where
\begin{displaymath}
\chi^{i}_{j} = \frac{\phi_{j}}{\phi_{i}}\quad i=1..3 \quad j=1..3.
\end{displaymath}
Here $i$ labels which chart is being used, and $j$ labels the components
of $\vec\chi$. $\vec\chi$ appears to have 3 components in this definition
but, as the component $\chi^{i}_{i}$ is always 1, $\vec\chi$ has only 2 degrees
of freedom. The chart labeled $i$ is ill defined on the equator $\phi_{i} = 0$,
and so all three charts are required to cover $\mathbb{R}P^{2}$ -- the first covers all
of the manifold except for an equator, the second all of this equator except
for 2 antipodal points and the third those two points.

A neat way to combine all three $\vec\chi$ charts is to use the matrix
$P_{ij}(x) = \phi_{i}(x)\phi_{j}(x)$. Each chart is easily extracted as
$\chi_{j}^{i} =\frac{P_{ij}}{P_{ii}}$ with no sum implied by repeated
indices.

\subsubsection{Topology}
\label{top}

The $n$th homotopy group of $\mathbb{R}P^{2}$ can be found quickly for
$n\ge2$ as $S^{2}$ is a universal covering space of $\mathbb{R}P^{2}$. A
theorem found in \cite{croom} states that this implies that $\pi_{n}(S^{2})
\cong \pi_{n}(\mathbb{R}P^{2}) \quad \forall \quad n \ge 2$. $\pi_{1}(\mathbb{R}P^{2})$
may be found to be the second symmetric group, $\mathbb{Z}_{2}$. To
summarise:
\begin{align} \label{pione}
\pi_{1}(S^{2}) \cong e \quad \pi_{1}(\mathbb{R}P^{2}) \cong \mathbb{Z}_{2}, \\
\label{pitwo}
\pi_{2}(S^{2}) \cong \mathbb{Z} \quad \pi_{2}(\mathbb{R}P^{2}) \cong \mathbb{Z}, \\
\label{pithr}
\pi_{3}(S^{2}) \cong \mathbb{Z} \quad \pi_{3}(\mathbb{R}P^{2}) \cong \mathbb{Z},
\end{align}

\subsubsection{Winding Number}

The class of field configuration $\vec\phi$ may be characterised by the degree of
map, or winding number. To calculate the winding number we must calculate the
pullback of the two form on the target space; the map $\vec\phi$ from the physical
space to the target space induces a natural map (the ``pullback'') from the two
form on the target space to a two form on the physical space. If the two form
on the physical space is then integrated over all space the result must be a multiple
of an integer. By renormalising we get a formula for the integer valued winding number:
\begin{equation}
T = \frac{1}{8\pi}\int d^{2}x \, \epsilon_{ij} \epsilon_{abc} \phi_{a} (\partial_{i}\phi_{b})
(\partial_{j}\phi_{c}).
\end{equation}
It should be noted that the transformation $\vec\phi \mapsto -\vec\phi$ leads
to $T \mapsto -T$. This is because $\mathbb{R}P^{2}$ is a non-orientable manifold
-- a two form is not well defined on $\mathbb{R}P^{2}$ as its sense is changed
by a translation along a non-trivial loop. In fact it may be shown \cite{cond} that
field configurations of $\vec\phi$ are characterised not by $\pi_{2}(\mathcal{M})$
but by $\pi_{2}(\mathcal{M})/\pi_{1}(\mathcal{M})$. In the case of an
$\mathbb{R}P^{2}$ target space this means that field configurations are
characterised by $\mathbb{Z}/\mathbb{Z}_{2}$.

\subsection{Numerical Techniques}

The models outlined above are highly nonlinear and more or less completely
intractable analytically. The models are, however, open to study using numerical
techniques. One may write a simulation on a discrete grid and analyse the evolution
of an arbitrary initial state. With a clever ansatz for the initial conditions,
with or without the use of a relaxation routine, the behaviour of minimal enery
field configurations may be studied.

We simulated the sigma and Skyrme models numerically, using $\mathbb{R}P^{2}$ as
the target space. Our simulations were based on a 200 $\times$ 200 point grid using
the nine point laplacian and derivatives outlined in \cite{bernard}, together with a
fourth order Runge-Kutta algorithm for simulating time evolution. The timestep
length was set to half the gridpoint spacing. To give the $\mathbb{R}P^{2}$ topology
the state vector was stored using the $P$ matrix described in section (\ref{geom})
for position on the target space, and $Q_{ij}=\dot{P}_{ij}=\phi_{i}\dot{\phi}_{j} + 
\dot{\phi}_{i}\phi_{j}$ for the rate of change of position on the target space with
time, where a dot is used to denote derivatives with respect to time.
The simulation then evaluated derivatives by determining the field as the equivalent
$\vec\phi$ for any given set of nine points, with the central point mapped to
the north pole of $S^{2}$. Reflections from the boundaries of the grid were reduced
by damping out kinetic energy in a region near the boundary. The fields were kept
on manifold by applying the following transformations each timestep:
\begin{align}
P'_{ij} & = \frac{P_{ij}}{P_{kk}}, \\
Q'_{ij} & = \frac{Q_{ij} - P'_{ij}Q_{kk}}{\sqrt{P_{kk}}}.
\end{align}
These transformations ensure that $Tr(P) = 1$ (i.e. the field lies on $\mathbb{R}P^{2}$)
and that $Tr(Q) = 0$ (i.e. the rate of change of the field is tangential
to the surface of the field manifold).

\section{The $\mathbb{R}P^{2}$ Valued New Baby Skyrme Model}

Simulations of this model were in close qualitative and quantitative agreement
with simulations using an $S^{2}$ target space.

\subsection{The Hedgehog Anzatz}

As in the $S^{2}$ model the simplest ansatz for a skyrmion is the radially symmetric
ansatz, usually refered to as the `hedgehog' ansatz \cite{tom}. For a skyrmion of
topological charge one this ansatz gives:
\begin{equation} \label{hedge}
\vec \phi = \left(
\begin{array}{c}
\sin[f(r)]\cos(\theta - \gamma)  \\ \sin[f(r)]\sin(\theta - \gamma) \\ \cos[f(r)]
\end{array} \right).
\end{equation}
Here $f(r)$ is known as the profile function and $\gamma$ is an arbitrary phase
parameter. This profile function is arbitrary
up to the boundary condition that $f(0) = (2n+1)\pi$ and $f(\infty) = 2m\pi$
where $n,m\in \mathbb{Z}$. To minimise the energy of the ansatz one may determine the
energy of the field as a functional of the profile function and its first derivative,
as shown below:
\begin{equation} \label{enfunc}
E = 2\pi \int_{0}^{\infty}rdr\big(f^{'2} + \frac{\sin^{2}f}{r^{2}}(1+2\theta_{1}f^{'2})
 + \theta_{2}V(f)\big).
\end{equation}
One may then use the calculus of variations to extremise this energy with the appropriate
limits at $r = 0$ and $r = \infty$. The resulting Euler-Lagrange equation is a non linear 
second order ODE for the profile function. We solved this numerically, using the shooting method.
This ansatz leads to a stable single skyrmion. Note that our ansatz is identical to that of
the $S^{2}$ skyrmion, and so our result agrees with the result of Weidig \cite{tom}.
Simulations showed this ansatz to be stable, again in agreement with Weidig.
Note that the energy has been defined such that the energy of a hedgehog ansatz
soliton with $\theta_{1}=\theta_{2}=0$ is given by $E=1$.

\subsection{Two Skyrmions}

As for the $S^{2}$ model, a two skyrmion ansatz may be arrived at by taking the stereographic
projection from the north pole of a one skyrmion:
\begin{equation}
W = \frac{\phi_{1} + i\phi_{2}}{1 - \phi_{3}}.
\end{equation}
This may then be combined with another skyrmion using the ansatz
\begin{equation}
\frac{1}{W_{T}}=\frac{1}{W_{1}} + \frac{1}{W_{2}},
\end{equation}
so that the final field $W_{T}$ is approximately equal to the field for any
skyrmion near that skyrmion (where $W = 0$) and the final field takes the vacuum value
of $W = \infty$ at spatial infinity. Using this ansatz we were able to reproduce the
attractive and repulsive channels found in \cite{tom}, as well as the ninety degree
scattering found in the attractive channel.

\subsection{Equivalence of $S^{2}$ and $\mathbb{R}P^{2}$ models for Smooth Maps}

In fact the two models are exactly equivalent as may be seen from the following argument.
Consider maps $f$, $\phi$:
\begin{equation} \label{itos}
f:I^{n}\mapsto S^{m},
\end{equation}
where $n<m$, and
\begin{equation} \label{stom}
\phi:S^{m}\mapsto M.
\end{equation}
\begin{figure}[ht]
\begin{center}
\includegraphics{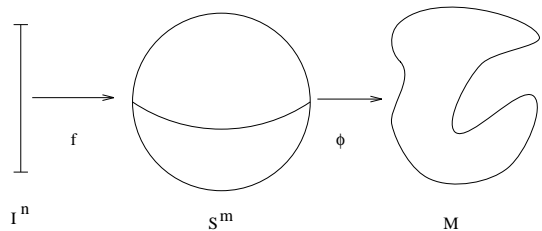}
\end{center}
\caption{Maps from $I_{n}$ to $S_{m}$ to $M$}
\end{figure}
As $\pi_{n}(S^{m})\cong e \quad \forall \quad n<m \quad f$ may always be smoothly
deformed to map all of $I^{n}$ to a point. This implies that, for all smooth maps $\phi$, all $n$-loops
in $S^{m}$ shall be the pre-image of an $n$-loop in $M$ corresponding to the
identity element of $\pi_{n}(M)$.

One consequence of this is that for $M \cong \mathbb{R}P^{2}$, $n=1$ and $m=2$,
all maps $\phi$ may be described as maps to $M \cong S^{2}$, as $\mathbb{R}P^{2}$
is universally covered by $S^{2}$, so the $\mathbb{R}P^{2}$ new baby Skyrme model
is identical to the $S^{2}$ baby Skyrme model if $\vec \phi$ is well defined at all
points.

It is also worthy of note that this argument suggests that simulations of
Fadeev-Hopf solitons \cite{fadknot,knot} using $\mathbb{R}P^{2}$ as a target space should reproduce the
simulation of such solitons with an $S^{2}$ target space unless defects are present.

\section{The Sigma Model with Defects}

The implication of the argument put forward above is that to discover new behaviour
in an $\mathbb{R}P^{2}$ model we must introduce a discontinuity into the field.
For such a point to be stable a circle around the discontinuity in physical space
must map to a non-trivial loop in the target space. If this is not the case then
the discontinuity can be removed by a continuous variation in the field. These
(point) discontinuities are analogous to disclination lines in a liquid crystal.
We shall refer to them as ``defects''. One consequence of having an odd number of
defects in the system is that a contour around the edge of the physical space must
map to a non-trivial loop on $\mathbb{R}P^{2}$. This means that it is no longer
possible to have the field tend to a single value at spatial infinity. As a result, it
is no longer easy to use a conventional potential term such as (\ref{newpot}) in
the Lagrangian. We may use the $\mathbb{R}P^{2}$ sigma model, so our model has
neither Skyrme nor a potential terms, or we may use an unconventional potential term.
One possible potential term would be of the form
\begin{equation} \label{easypl}
\mathcal{L}_{V}=\frac{1}{4} (\vec\phi\cdot\vec\phi_{mass})^{2}),
\end{equation}
which we shall call the ``easy plane'' potential. This effectively gives a mass to one
of the fields ($\phi_{mass}$) where the potential (\ref{newpot}) gives mass to
both fields orthogonal to $\phi_{vac}$. Easy plane baby skyrmions have not previously
been studied and so are worthy of attention in their own right. We have confined
our work to the sigma model in all that follows.

It should be noted that if there is no single value for the field at infinity
we are no longer compactifying the physical space to $S^{2}$. As a result winding
number is no longer necessarily conserved. In fact, the physical space is topologically
$S^{1}\times\mathbb{R}^{1}$ as not only is the space not compactified at infinity, it
also has no well defined map at the point of the defect. Whilst the mapping from the
target space is frozen at infinity, it is not frozen around the defect. The circle
around the defect necessarily maps to a non-trivial curve in $\mathbb{R}P^{2}$. If
we represent $\mathbb{R}P^{2}$ as a sphere, where antipodal identification is allowed,
a non-trivial curve is a line between a pair of (arbitrary) poles. If we have a
defect-soliton system, one of these curves represents both the image of a circle at
infinity in the physical space, and the image of an infinitesimal circle around the
defect. As a circle is contracted in from infinity to the defect, the image curve
rotates around the sphere, hinged on the polar identification, wrapping the sphere
once. As the image of the infinitesimal circle around the defect is able to move, it may
rotate around the identified poles, unwinding the soliton like object.

We decided to study the behaviour of a soliton like lump in the presence of a point
defect. To do this we need an ansatz for a soliton-defect system to use as an initial
conditon for our simulation. Such an ansatz needs to be sufficiently close to equilibrium
to avoid large quantities of radiation perturbing the system -- if the system is far from
equilibrium much of the excess energy will be shed from the solitonic objects as radiation.

\subsection{Defect Ansatz}

The most obvious ansatz for a defect is a radially symmetric, planar field.
A defect at $r=0$ may be written in polar coordinates as
\begin{equation} \label{defans}
\vec \phi = \left(
\begin{array}{c}
\sin \frac{\theta}{2} \\ 0 \\ \cos \frac{\theta}{2}
\end{array} \right),
\end{equation}
which, using the stereographic projection from the south pole,
\begin{equation} \label{stenth}
W = \frac{\phi_{1} + i\phi_{2}}{1 + \phi_{3}},
\end{equation}
is given by:
\begin{equation}
W = \tan \frac{\theta}{4}.
\end{equation}
This ansatz is defined for $0\leq \theta < 2\pi$. On the line $\theta = 0$ there
appears, at a first glance, to be a discontinuity in the field, but as $\vec \phi$
is identified with $- \vec \phi$ there is no such discontinuity. In fact any
field configuration which contains a defect must have a line from the defect to infinity
along which the identification between $\vec\phi$ and $-\vec\phi$ is made. It is simple
to show analytically that the above field configuration is in equilibrium. Simulations
also show that this field configuration is stable. It should be noted that this field
configuration has an infinite action arising from a logarithmic divergence as $r$ tends
to infinity or as $r$ tends to zero. This is not a problem if the physical system we are
modeling is a liquid crystal. This is because such a system is, firstly, finite in
extent, and, secondly, a discrete system. The field approximation is clearly not
appropriate at $r = 0$. Care must be taken when simulating such an object numerically to
place such an object away from a grid point.

\subsection{`Glued' Ansatze}

\subsubsection{The Three Region Ansatz}

\begin{figure}[ht] 
\begin{center}
\includegraphics{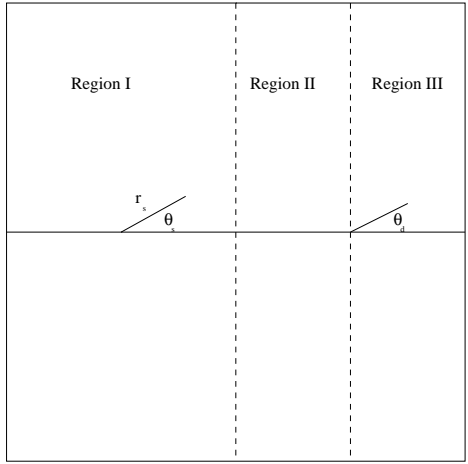}
\end{center}
\caption{Regions and Coordinates for Glued Ansatze}
\label{regions}
\end{figure}

One ansatz for a defect-soliton system which we tried early on involved `glueing'
together three regions in an attempt to graft together a known ansatz for the soliton
to a defect like object -- see figure (\ref{regions}).
In region I the field followed the standard lump ansatz for the O(3) sigma model.
This is given by the hedgehog ansatz (\ref{hedge}) with the profile function given by:
\begin{equation} \label{lump}
f(r)= \tan^{-1}\frac{2\lambda r}{1-\lambda^{2}r^{2}},
\end{equation}
where $\lambda$ is an arbitrary (real) parameter describing the (inverse) `width' of the lump.
In region II the field is constant and equal to the vacuum value of region I -- there should
be little discrepancy at the border between regions I and II as a result, and any variation may
be interpolated across several grid points in the simulation. In the region III the field
is given by:
\begin{equation}
\vec \phi = \left(
\begin{array}{c}
\cos\theta_{d} \\ 0 \\ \sin \theta_{d}
\end{array} \right),
\end{equation}
for $-\pi \le \theta_{d} < \pi$. The region II and region III fields match along
the line $\theta = \frac{\pi}{2}$ and, due to the identification $\vec\phi \equiv -\vec\phi$,
along the line $\theta = -\frac{\pi}{2}$.  This places a defect at the origin.
Unfortunately this defect is far from stable -- the energy of this defect is double
that of the defect described by (\ref{defans}). Thus in the simulations of this total field
configuration we have seen the emission of waves of radiation by the defect. These waves
affected the skyrmion. To reduce these problems we have modified our ansatz as
detailed below.

\subsubsection{The Two Region Ansatz}

\begin{figure}[ht]
\begin{center}
\includegraphics{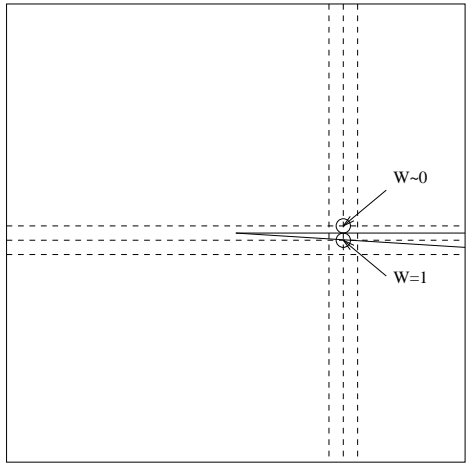}
\end{center}
\caption{Diagram showing discretisation problem}
\label{glued}
\end{figure}

We modified the above ansatz by replacing the field in regions II and III by the field below:
\begin{equation}
W = \tan \frac{\theta_{d}}{4} e^{-k(r_{d}-r_{0})},
\end{equation}
where $0 \le \theta_{d} < 2\pi$ and W is the stereographic projection from the south pole.
At a first glance this
field looks as though it might tend to the vacuum value in all directions whilst looking
like a defect at small radii. In fact, this is not the case -- the field is smooth away
from the origin, and therefore a loop around this field at infinity must be non-trivial.
The field must vary dramatically as $\theta$ approaches $2\pi$ and $r$ approaches infinity.
We had hoped that this would not be a problem in the (finite) region of the simulation,
but discretisation brought its own problems, as shown in figure (\ref{glued}). The line
where $W = 1$ must cross the last line of grid points on the large $\theta_{d}$ side of
the $\theta_{d} = 0$ line.
This leads to a point where the variation in the field from one grid point to
the next is as large as it can be, resulting in a region of high energy density.
In fact this region also has defect type winding number -- the ansatz has created
a second defect in a far from equilibrium configuration. In fact if one constructs
an ansatz which tends to constant field at spatial infinity (or the edge of the
simulation region) then the space must contain an even number of defects. A loop
at large r which maps to one point in $\mathbb{R}P^{2}$ must correspond to the
trivial element of $\pi_{2}(\mathbb{R}P^{2}) = \mathbb{Z}_{2}$, so the number
of defects inside this loop must be even.

\subsection{The Stereographic Defect - Soliton Ansatz}
\label{stanz}

To create an ansatz for a field configuration including both a soliton and a defect we
must re-express our field configurations using the stereographic map, $W$.
A projection from the south pole causes the north pole, $\phi_{3} = 1$, to map to $W = 0$
and the south pole to map to $W = \infty$. Note that the antipodal identification under
this map is given by
\begin{equation} \label{wident}
W \mapsto -\frac{1}{W^{*}}.
\end{equation}
Our ansatz must a) look like a soliton near the soliton, b) look like a defect near the
defect, c) look like a defect at infinity and d) be smooth at all points away from the defect. 

A single soliton at the origin may be described by the field
\begin{equation} \label{phisol}
\vec \phi = \left(
\begin{array}{c}
\alpha \cos(\theta-\gamma) \\ \alpha \sin (\theta-\gamma) \\ \beta
\end{array} \right),
\end{equation}
where $\alpha = \frac{2\lambda r}{1+\lambda^{2}r^{2}}$,
$\beta = \frac{1-\lambda^{2}r^{2}}{1+\lambda^{2}r^{2}}$, $\gamma$ is an arbitrary (real)
phase parameter, $\lambda$ is an arbitrary (real) parameter characterising the width of the
soliton, and $(r,\theta)$ are polar coordinates. Note that this is the hedgehog ansatz
with the profile function given by (\ref{lump}). To combine this with a defect we 
choose the point of projection of both fields such that the soliton at infinity maps to a
number of modulus $1$, and the line along which the defect field identifies approaches
$0$ from one side and $\infty$ from the other. These fields are given below.
\begin{equation} \label{wsol}
W_{s} = \frac {\alpha \sin (\theta - \gamma) + i\beta}{1+\alpha \cos (\theta - \gamma)},
\end{equation}
\begin{align} \label{wdef}
W_{d} & = \frac {\sin(\frac{\theta}{2})}{1 + \cos(\frac{\theta}{2})}\nonumber\\
      & = \tan(\frac{\theta}{4}).
\end{align}
If we multiply these fields we find
\begin{equation} \label{wansa}
W_{T} = \frac {\alpha \sin (\theta_{s} - \gamma) + i\beta}{1+\alpha \cos (\theta_{s} - \gamma)}
\tan(\frac{\theta_{d}}{4}),
\end{equation}
where $\theta_{d}$ is the polar angle coordinate with respect to the position of the defect
and $(r,\theta_{s})$ are the polar coordinates with respect to the position of the soliton.
This ansatz obeys conditions c) and d) above. c) is satisfied as at infinity $W_{T}=-i \times
W_{d}$. This is acceptable as multiplying a field configuration by a number of unit modulus
is merely a rotation about the axis of projection. d) is also satisfied as the line of identification
in $W_{d}$ is preserved by the multiplication -- $0 \times W_{s} = 0$ and (crudely speaking)
$\infty \times W_{s} = \infty$. Condition a) is fulfilled if the soliton and defect are
sufficiently separated for the variation in $W_{d}$ to be small across the scale of the
soliton. Condition b), however, is only fulfulled if $(\theta_{s}-\gamma) = (2n+1)\frac{\pi}{2}$
where $n\in\mathbb{Z}$.

\subsection{The Inhomogeneous Defect - Soliton Ansatz}

An alternative ansatz involves using the inhomogeneous coordinates outlined in section
(\ref{geomtop}). If we define a complex number
\begin{equation}
\mathcal{W}=\frac{\phi_{1} + i \phi_{2}}{\phi_{3}}
\end{equation}
then we can express the soliton field as a complex number which tends to $0$ at $r$ tends
to $0$ and $\infty$. If we add this to the defect field then the result must satisfy
conditions c) and d) outlined in section (\ref{stanz}). c) is satisfied because the map used
is a map of $\mathbb{R}P^{2}$, not $S^{2}$ and so needs no line of identification for the
defect ansatz, whilst d) is obeyed as the soliton field vanishes at infinity, leaving only
the defect field. Explicitly, the fields look like this:
\begin{align}
\mathcal{W}_{s} = \frac{\alpha}{\beta}e^{i(\theta_{s} - \gamma)}, \\
\mathcal{W}_{d} = \tan(\frac{\theta_{d}}{2}), \\
\label{inhom}
\mathcal{W}_{T} = \frac{\alpha}{\beta}e^{i(\theta_{s} - \gamma)} + \tan(\frac{\theta_{d}}{2}).
\end{align}
This expression obeys a) if $\mathcal{W}_{d}$ is small around the position of the soliton.
Condition b) is only met if $\mathcal{W}_{s}$ is small in the region of the defect, and
in the example above condition b) is less well satisfied if $\mathcal{W}_{s}$ has an
imaginary component at the defect.

\subsection{Simulation methods}

For simulations of a defect-soliton system we introduced a conformal grid in a similar manner
to Leese et al in \cite{peyrard}. This involved changing the physical coordinates $(x,y)$ to $(x',y')$
such that
\begin{equation}
X' = \frac{X}{1 + |X|}.
\end{equation}
This allows a much larger area to be simulated. The grid was $199 \times 199$ points
with $dX' = 0.01$. This conformal grid was introduced to reduce boundary effects.
Again, the timestep length was set to half the gridpoint spacing.

Another alteration to the simulation for defect-soliton systems was the introduction 
of a period of relaxation before the simulation was allowed to run freely. $Q_{ij}$
was set to zero throughout the grid every $10$ timesteps for the first $100$ timesteps
to take the simulation as close as possible to its minimal energy before the simulation
proper began.

\subsection{Results}

The soliton-defect system displayed certain generic characteristics -- a ``spreading'' channel
and a ``spiking'' channel according to the phase of the soliton. In the spreading channel the
soliton would become broader and broader until it significantly overlapped the defect, at
which point it would unwind, and the energy would be released to infinity as radiation. In the
spiking channel the soliton would become more and more localised until the variation in
the field became numerically untenable. The defect would remain stationary in all simulations,
whilst the soliton would move only a very small distance, to the extent that variations in
the width of the soliton would be far more significant than any movement of the maxima of
the soliton.

\subsubsection{Stereographic Ansatz}
\label{steans}

We used the stereographic ansatz (equation(\ref{wansa})) as our initial condition with $\lambda = 2.5$, and the initial
soliton position relative to the defect at $\theta_{d}= \pi$, $r_{d}= 2.5$. We shall discuss the
results of simulations with $\gamma$ set to $\frac{\pi}{2}$, $-\frac{\pi}{2}$, $0$ and $-\frac{\pi}{4}$.
After the initial period of relaxation energy was conserved to better than $0.05\%$ in all of these simulations. The
period of relaxation ended at $t=0.5$, after which the simulation ran freely until $t=9.995$, unless
the simulation became untenable due to soliton spiking.

\begin{figure}[hp]
\includegraphics[scale=0.45]{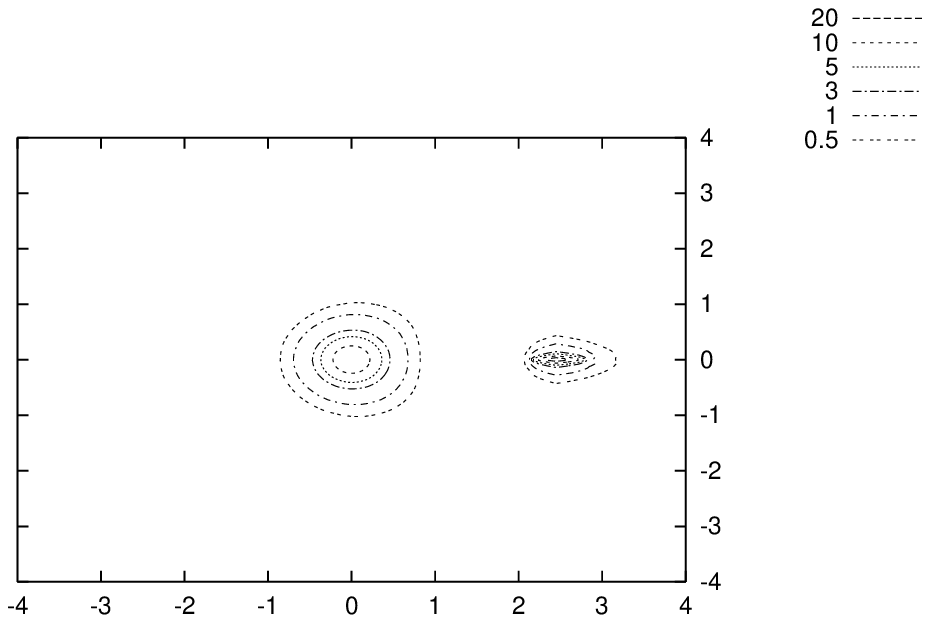}
\includegraphics[scale=0.45]{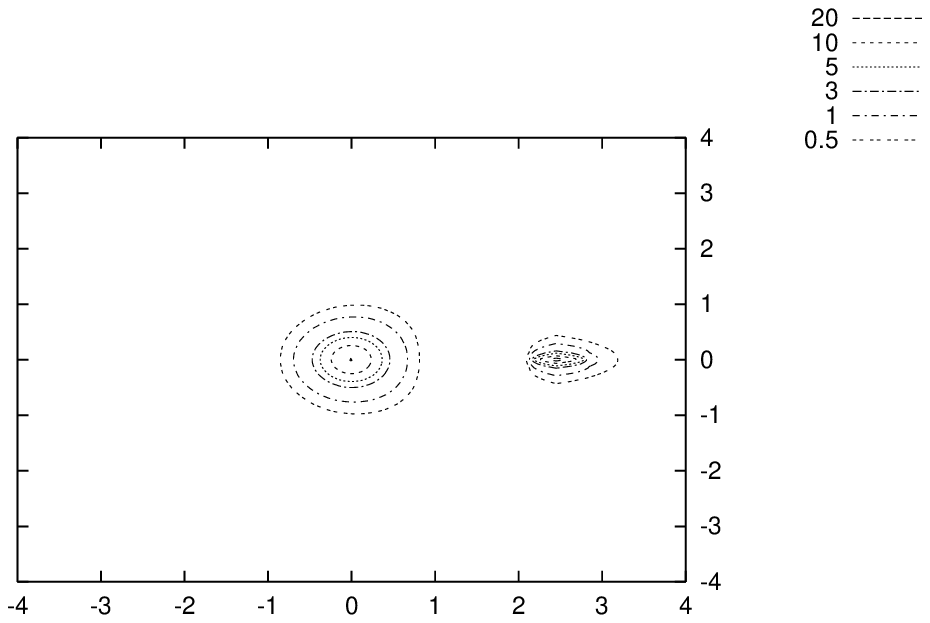}
\\
\includegraphics[scale=0.45]{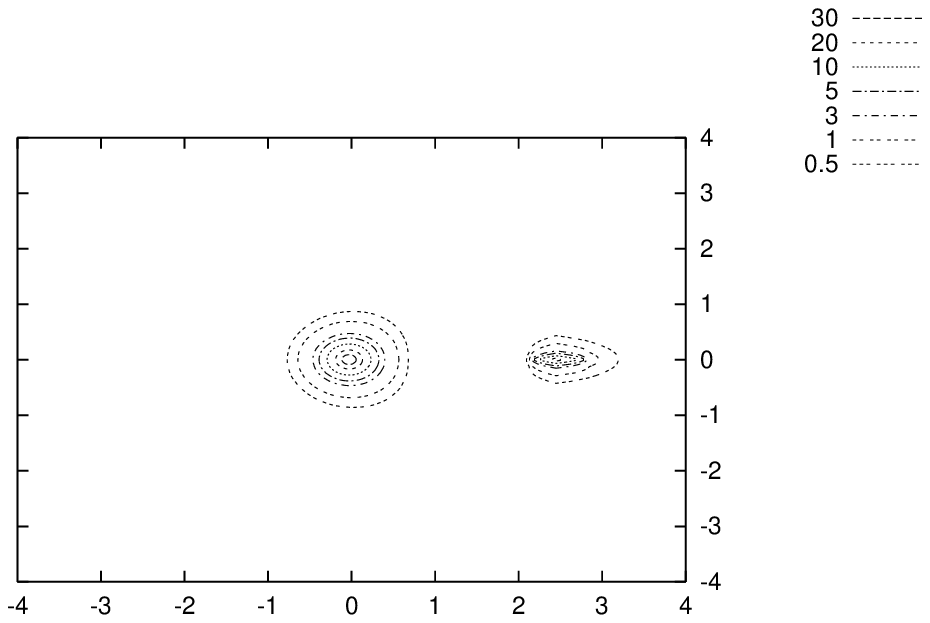}
\includegraphics[scale=0.45]{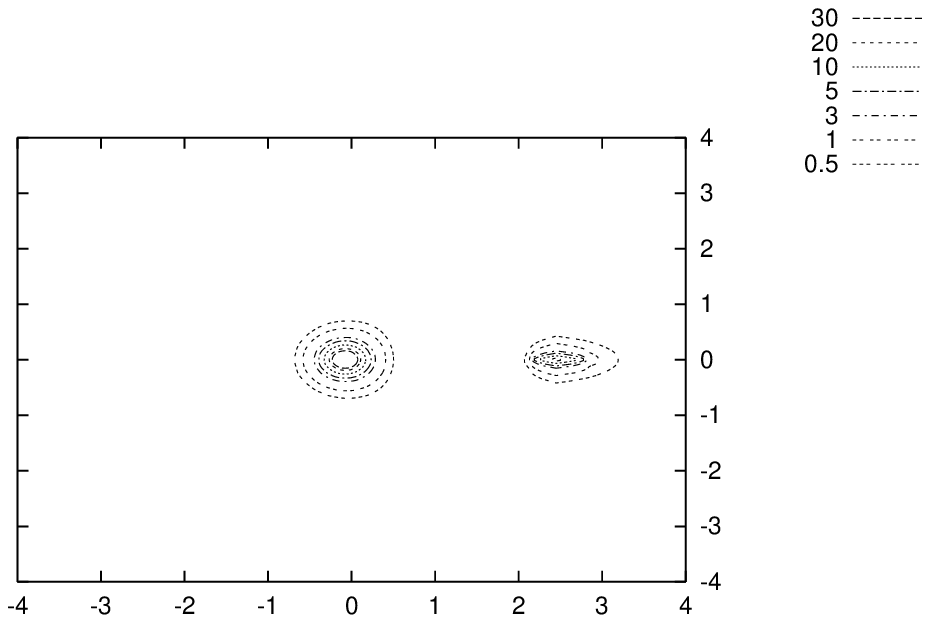}
\caption{Energy density for $\gamma = -\frac{\pi}{2}$ at t=0, 1, 2 and 3}
\label{spike}
\end{figure}

When the simulation was run with the phase initially set to $\gamma = -\frac{\pi}{2}$ the
soliton became narrower with time as described above (see figure (\ref{spike})). This
resulted in the simulation ending at $t=3.745$.

\begin{figure}[hp]
\includegraphics[scale=0.45]{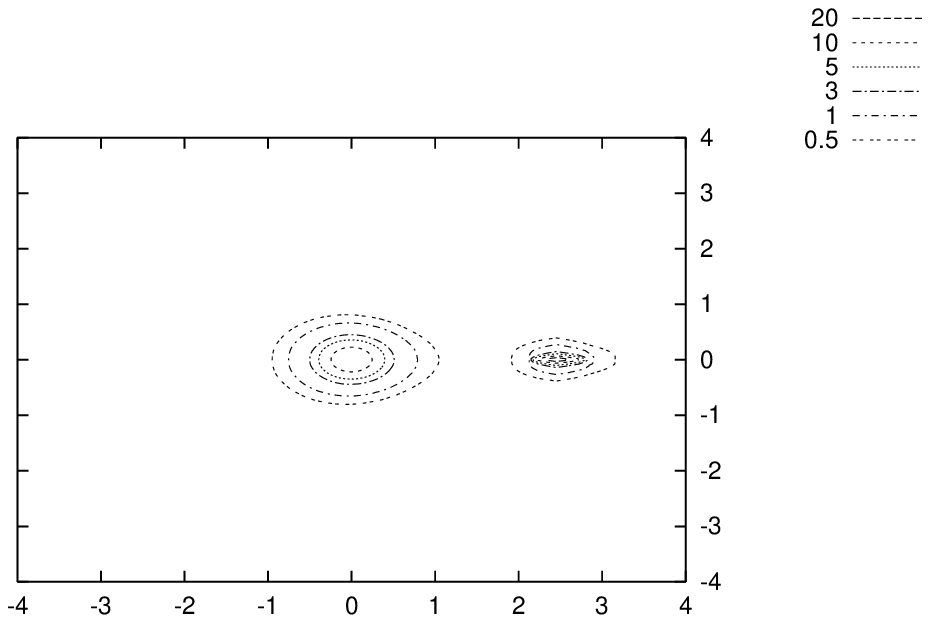}
\includegraphics[scale=0.45]{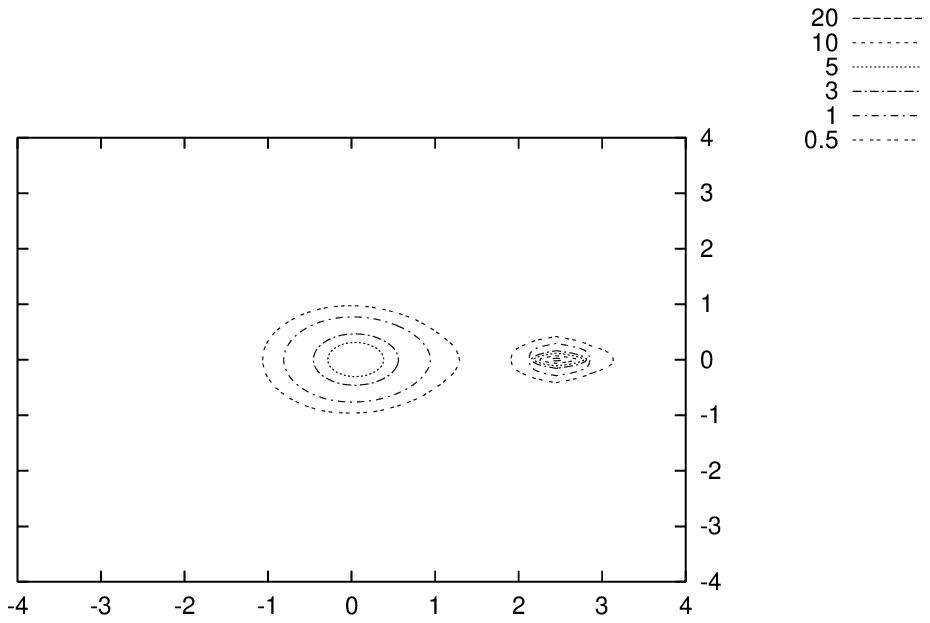}
\\
\includegraphics[scale=0.45]{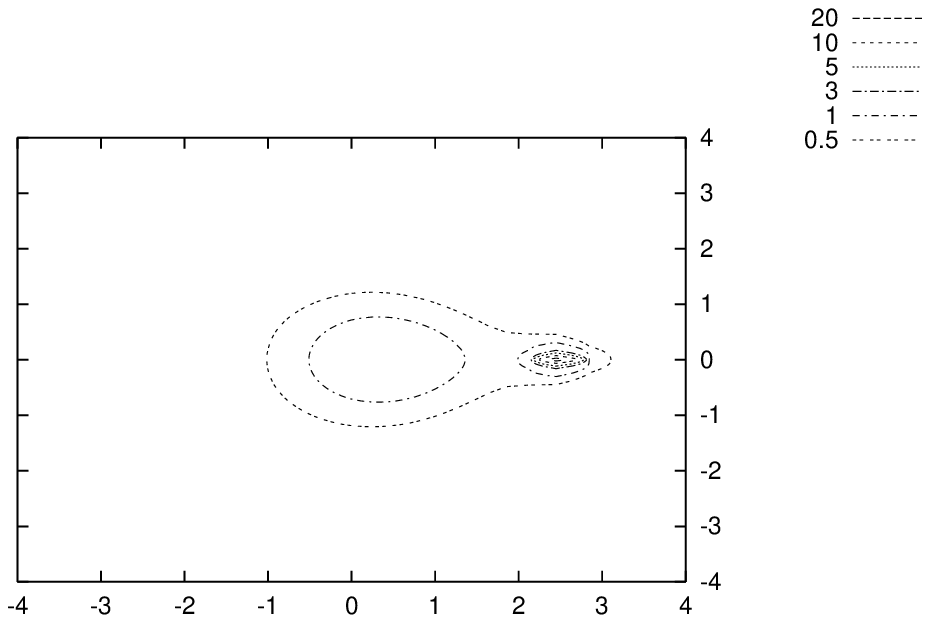}
\includegraphics[scale=0.45]{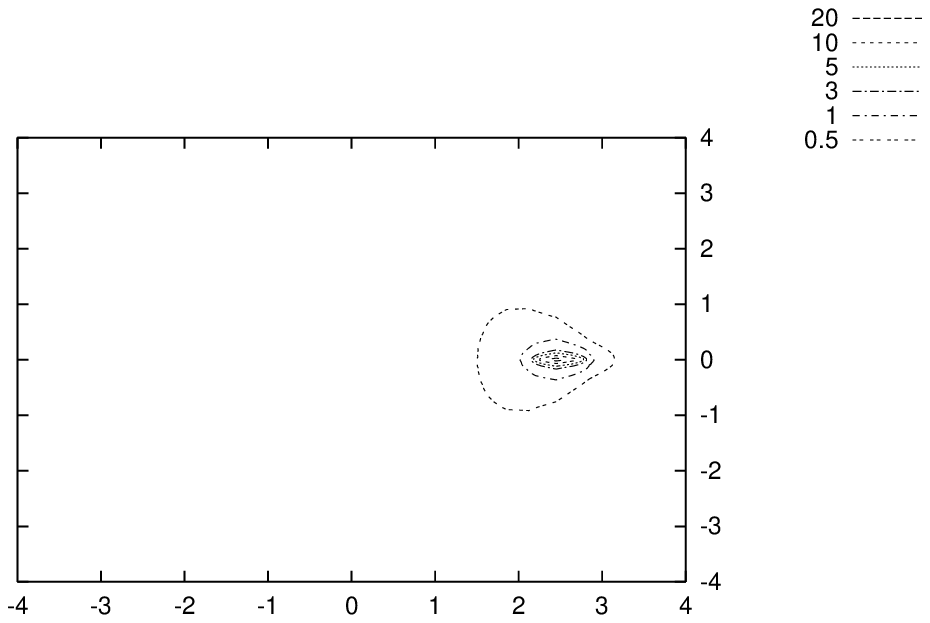}
\\
\includegraphics[scale=0.45]{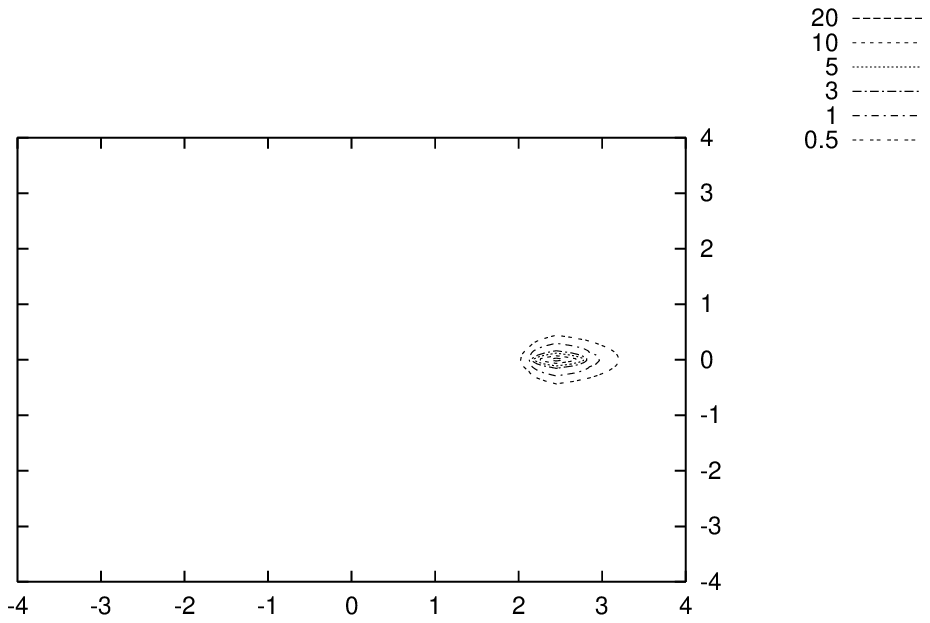}
\includegraphics[scale=0.45]{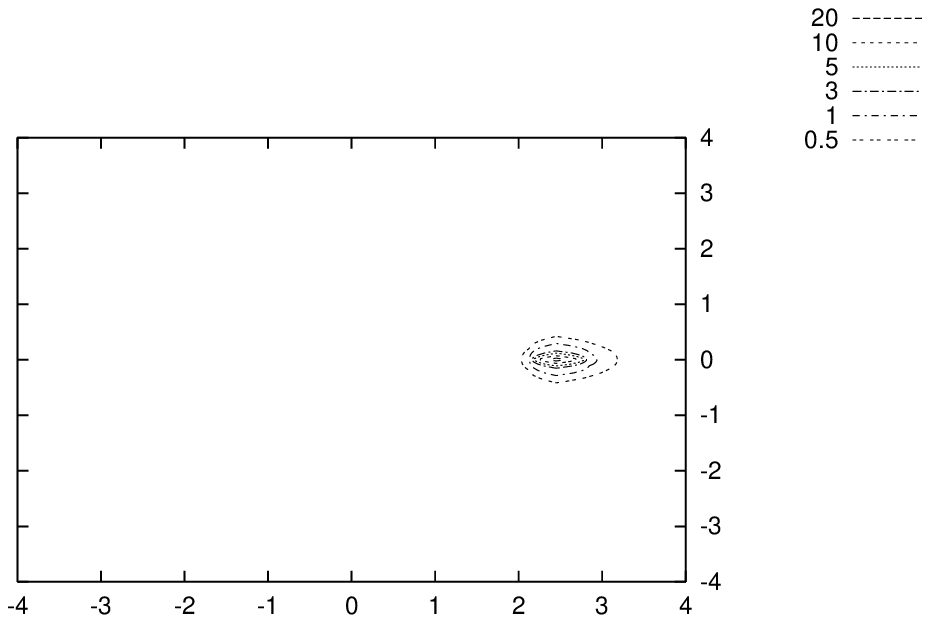}
\caption{Energy density for $\gamma = \frac{\pi}{2}$ at t=0, 2, 4, 6, 8 and 9.975}
\label{spread}
\end{figure}

When $\gamma = \frac{\pi}{2}$ the soliton spread out and eventually unwound as described above
(see figure (\ref{spread})). When $\gamma = 0$ was simulated the soliton was still subject to
spreading and unwinding -- the spreading channel is wider than the spiking channel.

\begin{figure}[hp]
\includegraphics[scale=0.45]{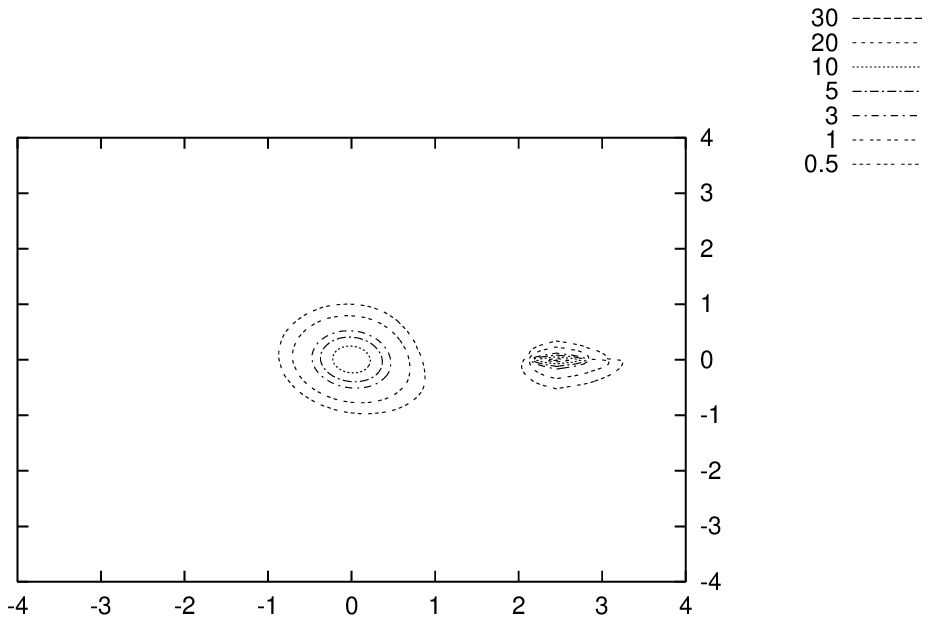}
\includegraphics[scale=0.45]{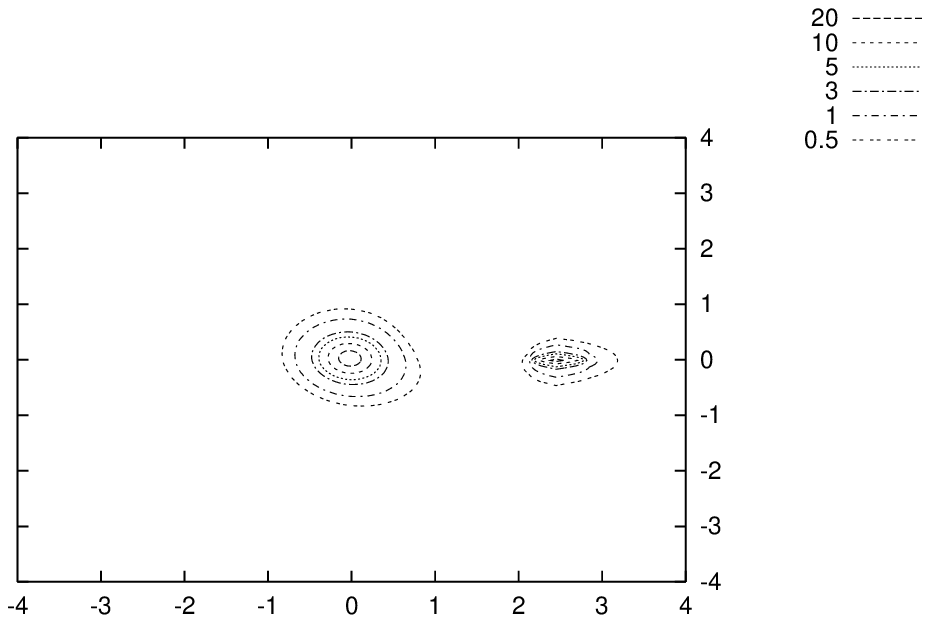}
\\
\includegraphics[scale=0.45]{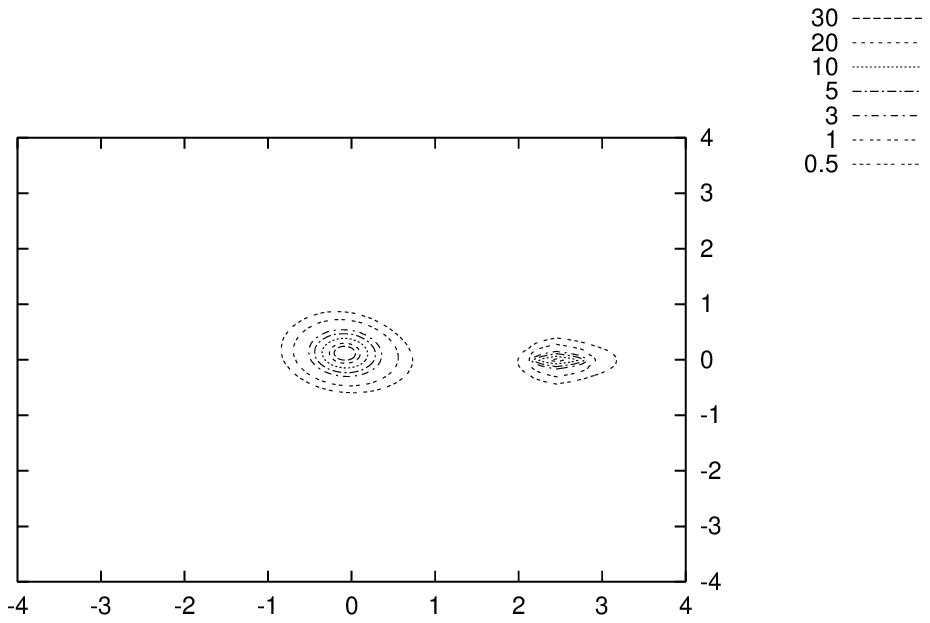}
\includegraphics[scale=0.45]{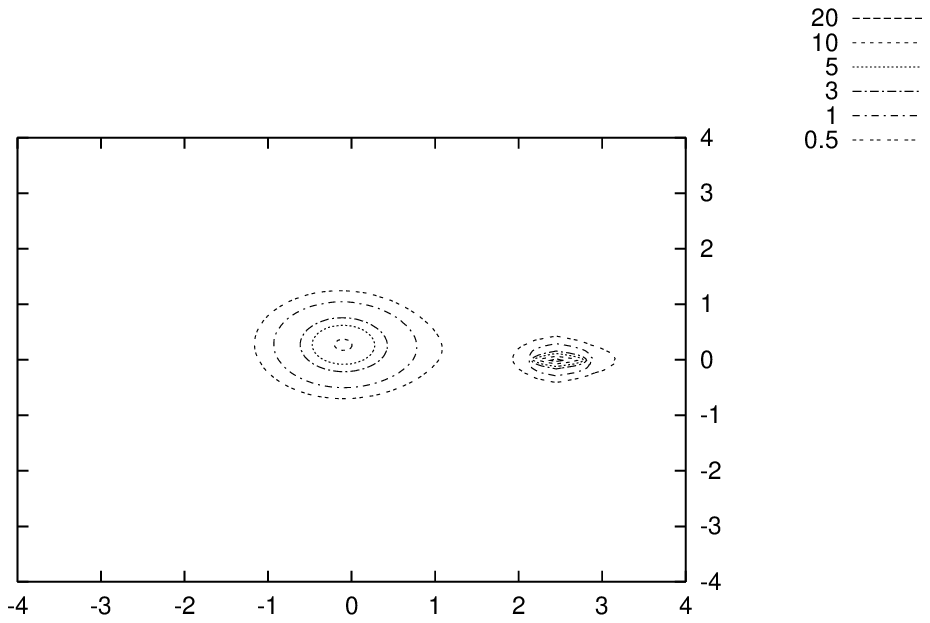}
\\
\includegraphics[scale=0.45]{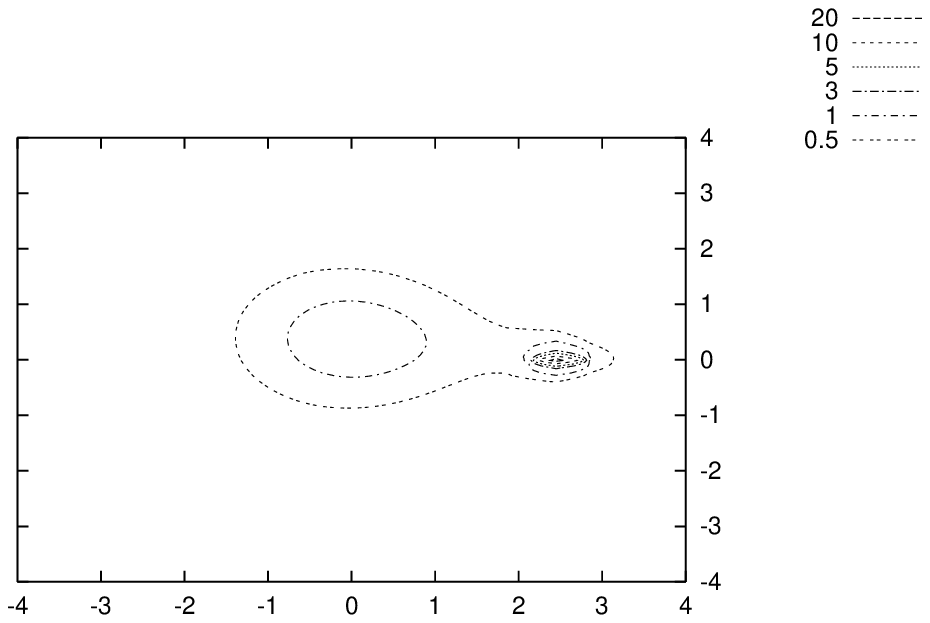}
\includegraphics[scale=0.45]{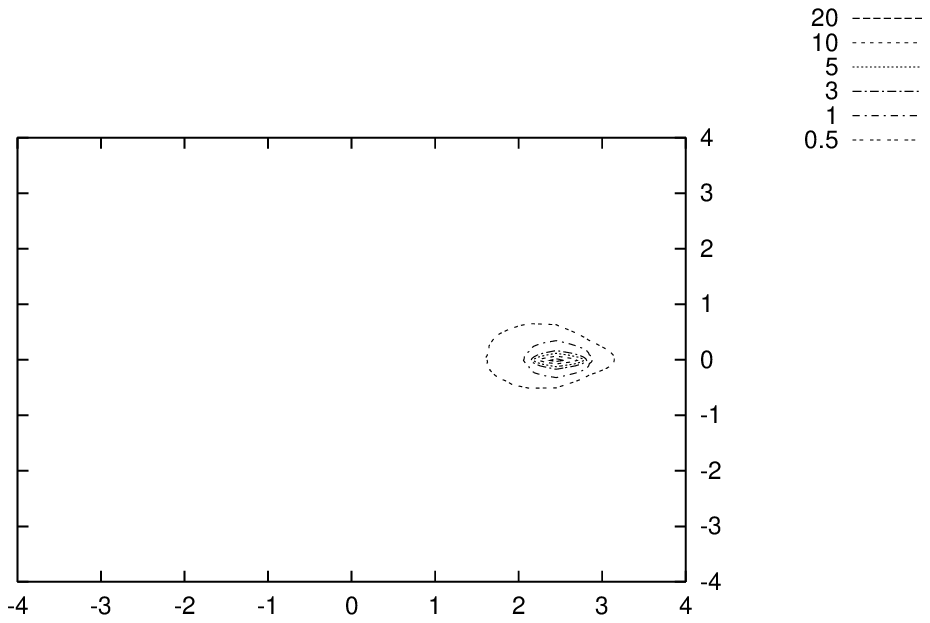}
\caption{Energy density for $\gamma = -\frac{\pi}{4}$ at t=0, 2, 4, 6, 8 and 9.75}
\label{spir}
\end{figure}

For $\gamma = -\frac{\pi}{4}$ the soliton began to spike, but after a period of time
the the rate of spiking slowed and the soliton began to spread, with the soliton eventually
unwinding (see figure(\ref{spir})). This also suggests that the spiking channel is unstable.

Table (\ref{penerg}) shows the energies of various simulations with initial conditions
given by this ansatz. Note that these energies depend heavily on the lattice spacing at
the defect, and are therefore only of value when comparing the different channels.
\begin{table}[h]
\begin{center}
\begin{tabular}{|c|c|c|c|c|}
\hline
$\gamma$     & Initial Energy & Energy after Relaxation & Time at end  & Final  \\
             &    ($t=0$)     &       ($t=0.5$)         & of Simulation &  Energy  \\
\hline
$\frac{\pi}{2}$ & 1.439 & 1.435 & 9.995 & 1.435 \\
\hline
0 & 1.520 & 1.512 & 9.995 & 1.512 \\
\hline
$-\frac{\pi}{2}$ & 1.595 & 1.590 & 3.745 & 1.591 \\
\hline
\end{tabular}
\caption{Energy of System in Stereographic Ansatz}
\label{penerg}
\end{center}
\end{table}

\clearpage

\subsubsection{Inhomogeneous Ansatz}
\label{inhans}

\begin{figure}[hbp]
\includegraphics[scale=0.45]{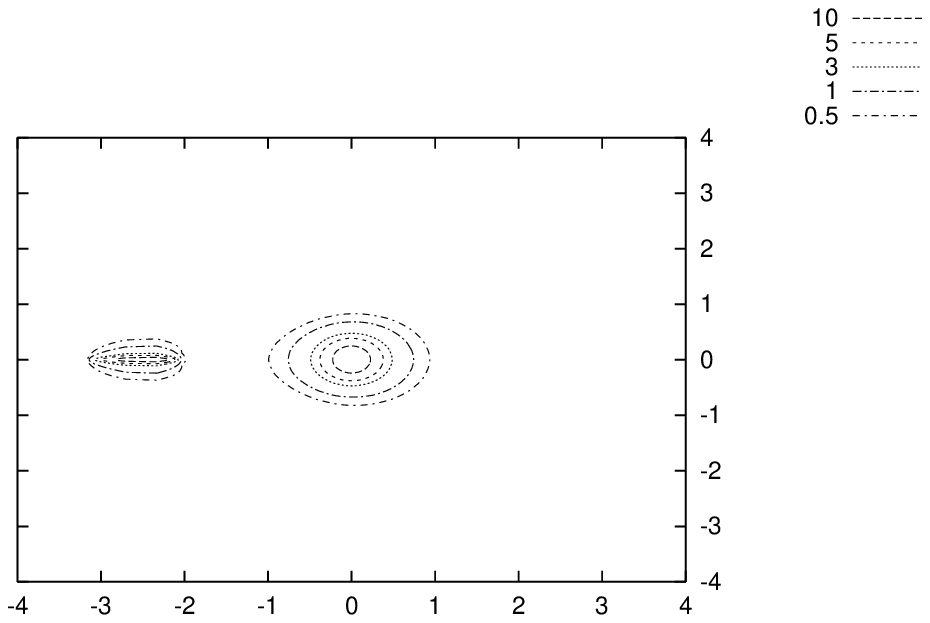}
\includegraphics[scale=0.45]{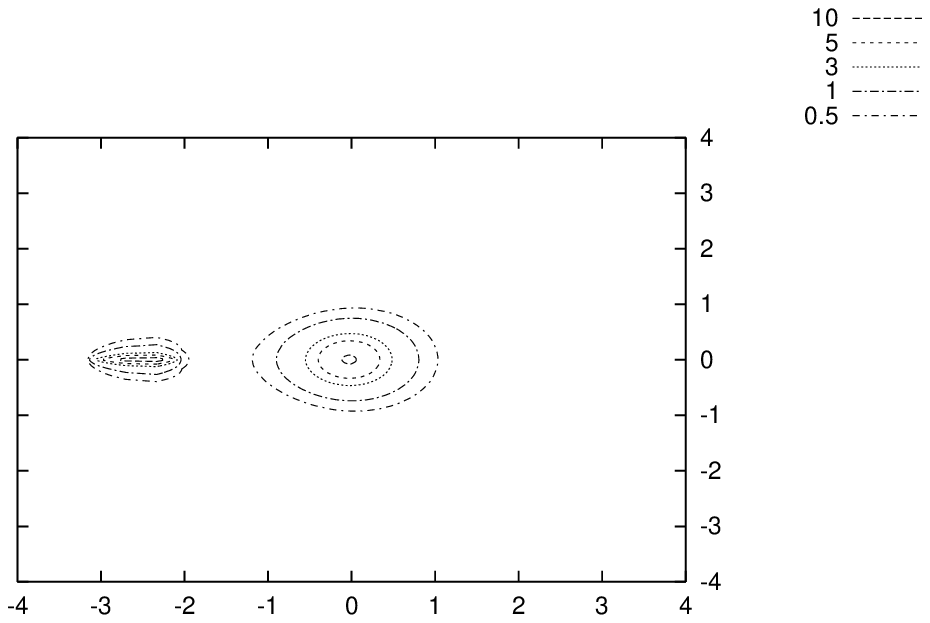}
\\
\includegraphics[scale=0.45]{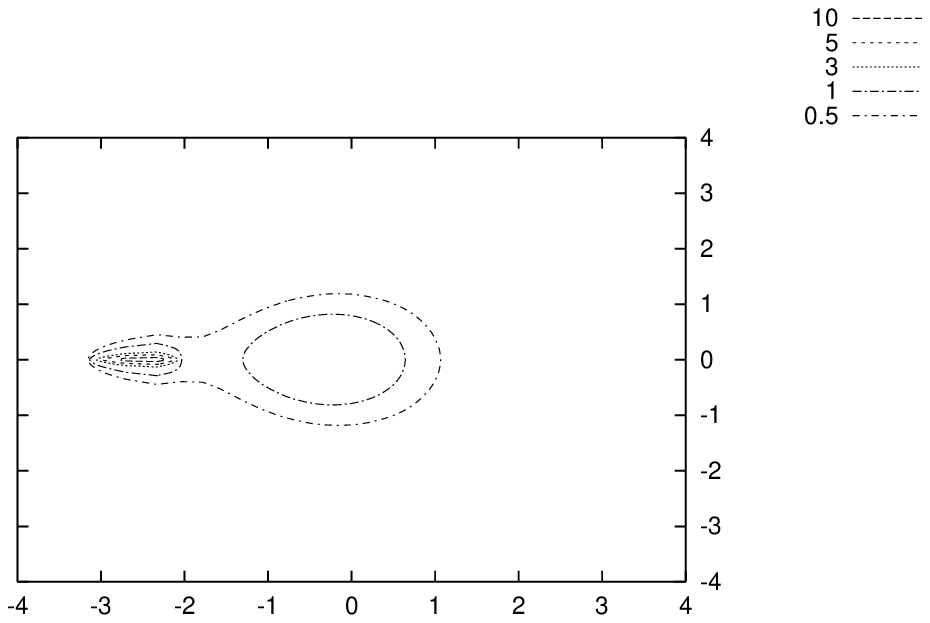}
\includegraphics[scale=0.45]{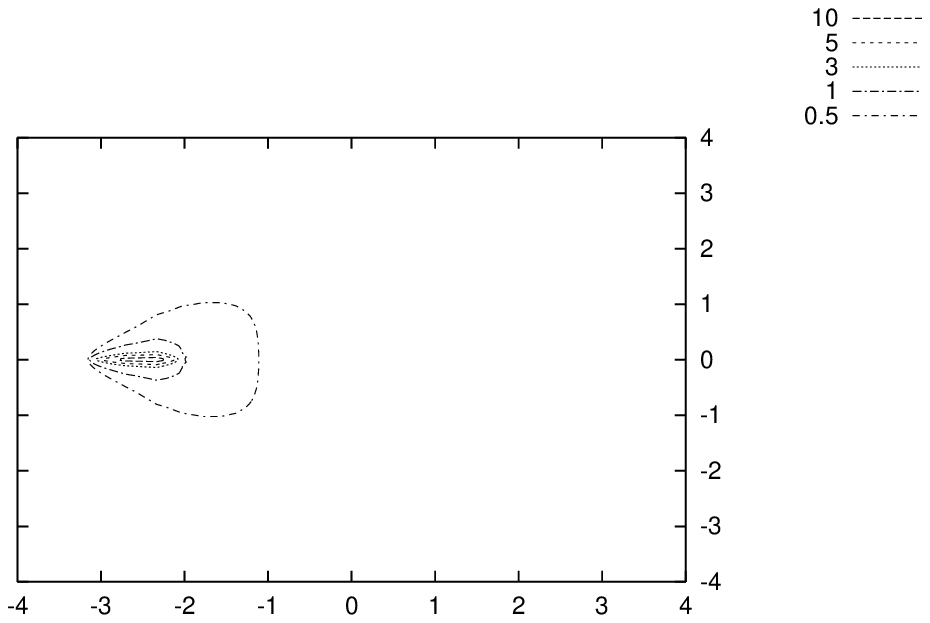}
\\
\includegraphics[scale=0.45]{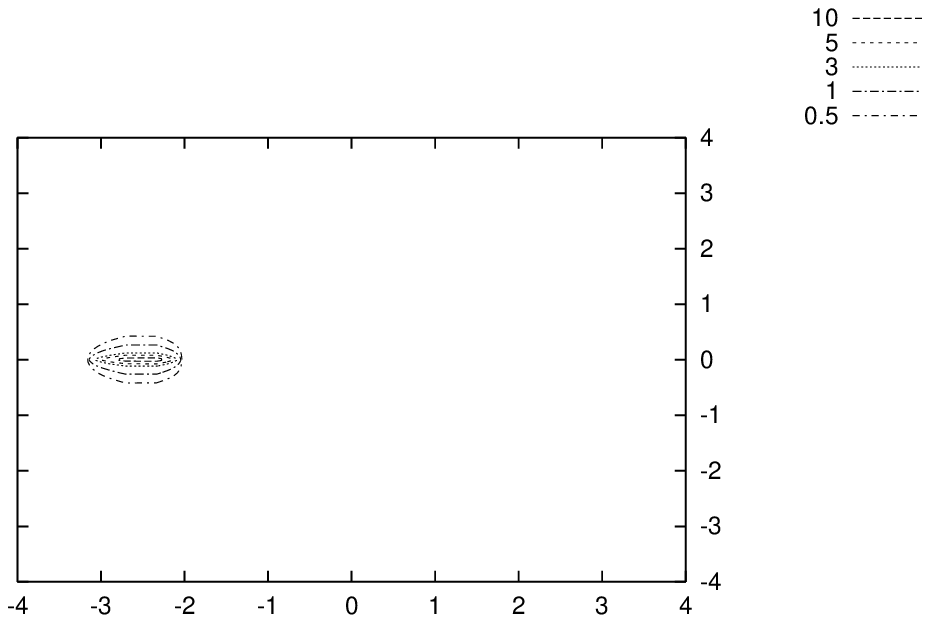}
\includegraphics[scale=0.45]{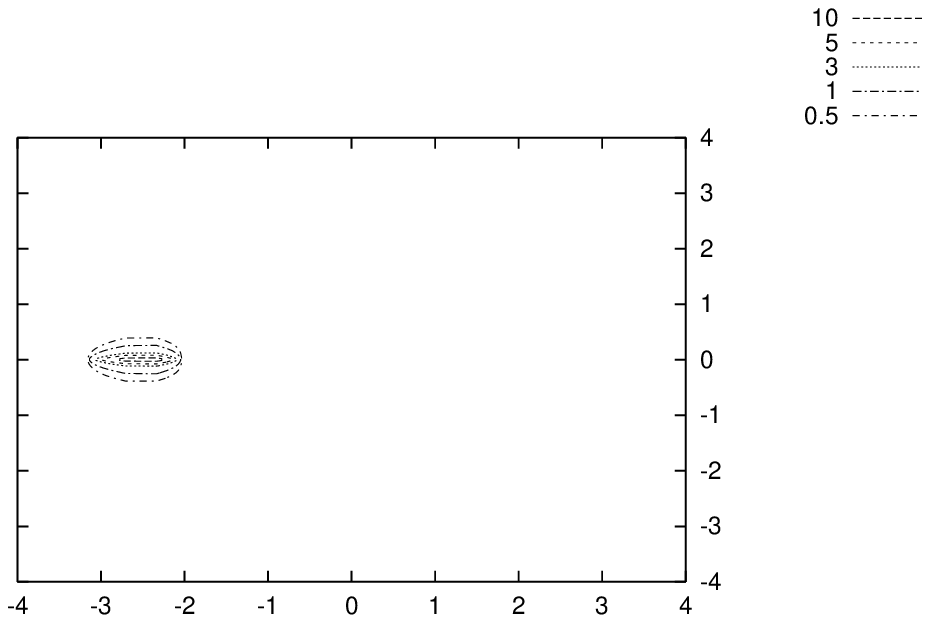}
\caption{Energy density for $\gamma = -\frac{\pi}{2}$ at t=0, 2, 4, 6, 8 and 9.975}
\label{sspread}
\end{figure}

\begin{figure}[hp]
\includegraphics[scale=0.45]{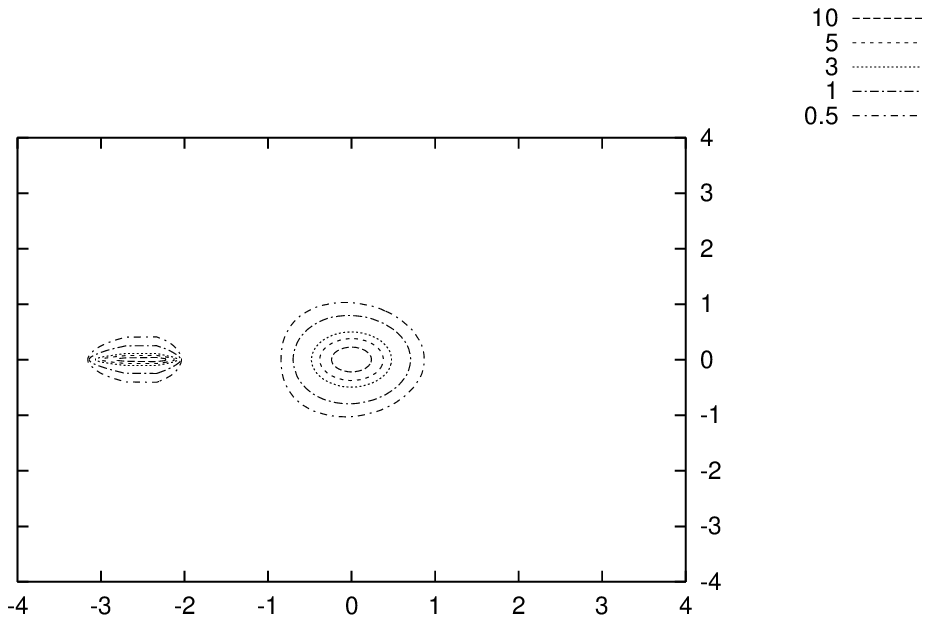}
\includegraphics[scale=0.45]{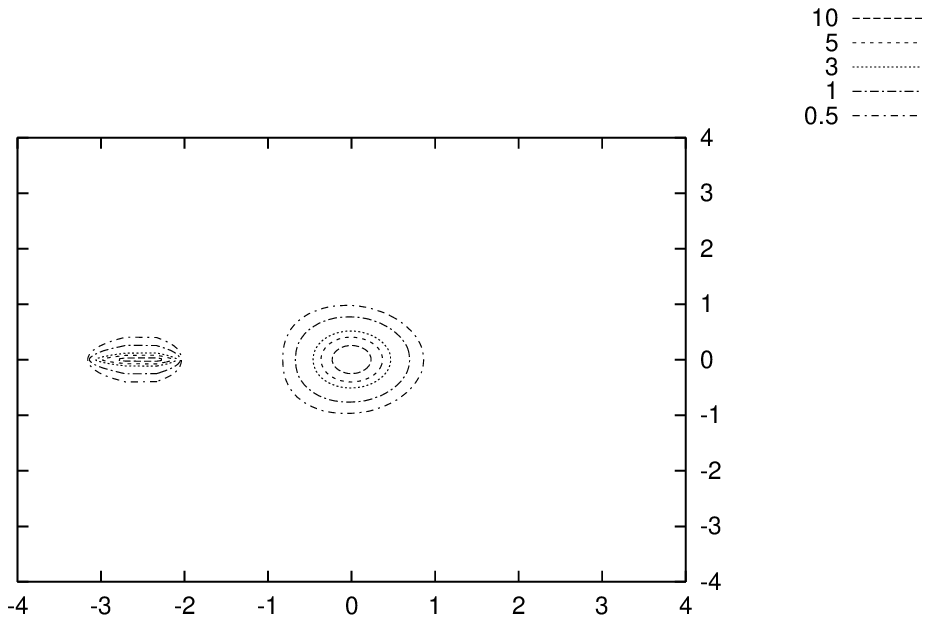}
\\
\includegraphics[scale=0.45]{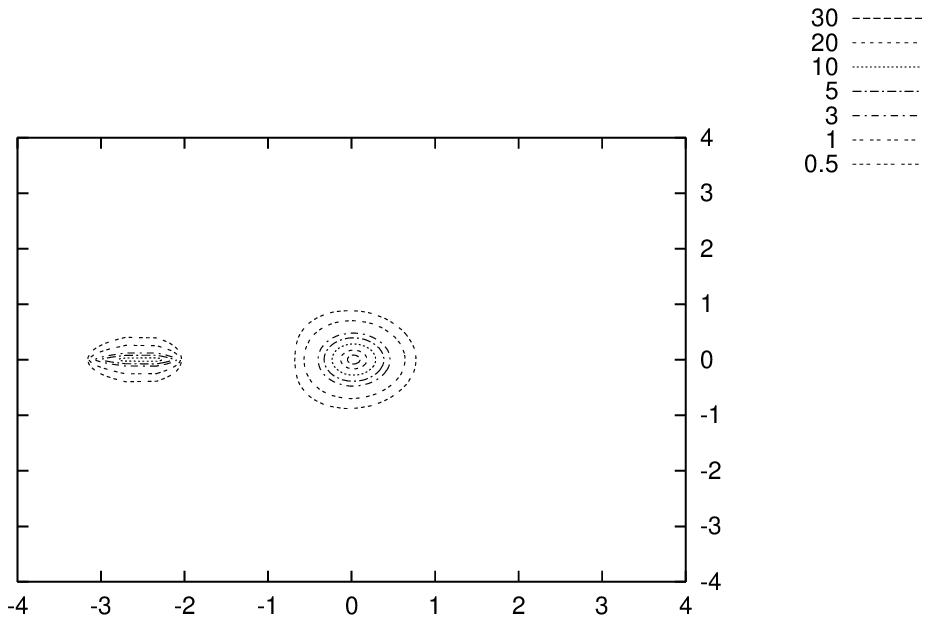}
\includegraphics[scale=0.45]{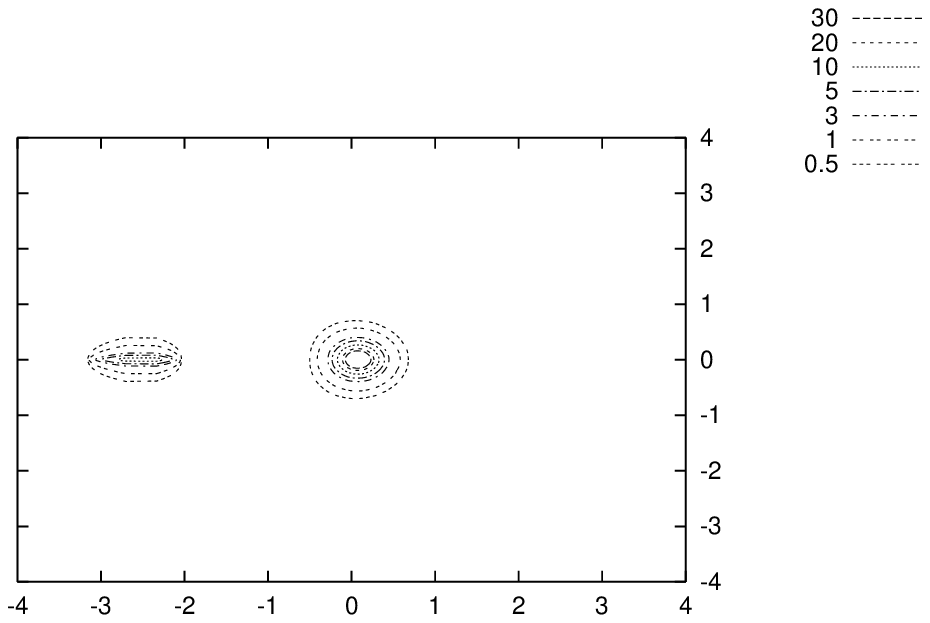}
\caption{Energy density for $\gamma = \frac{\pi}{2}$ at t=0, 1, 2 and 3}
\label{sspike}
\end{figure}

\begin{figure}[hp]
\includegraphics[scale=0.45]{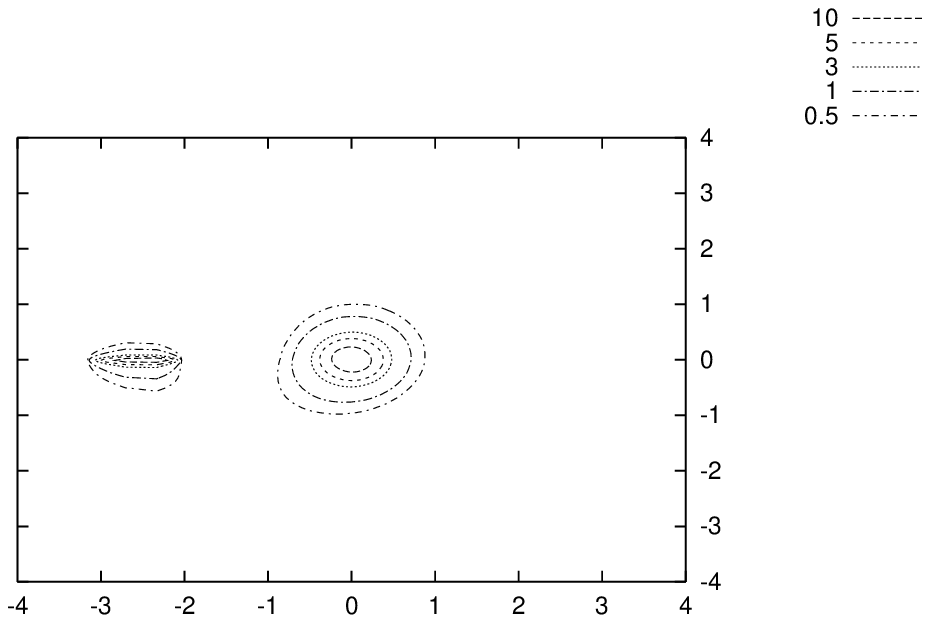}
\includegraphics[scale=0.45]{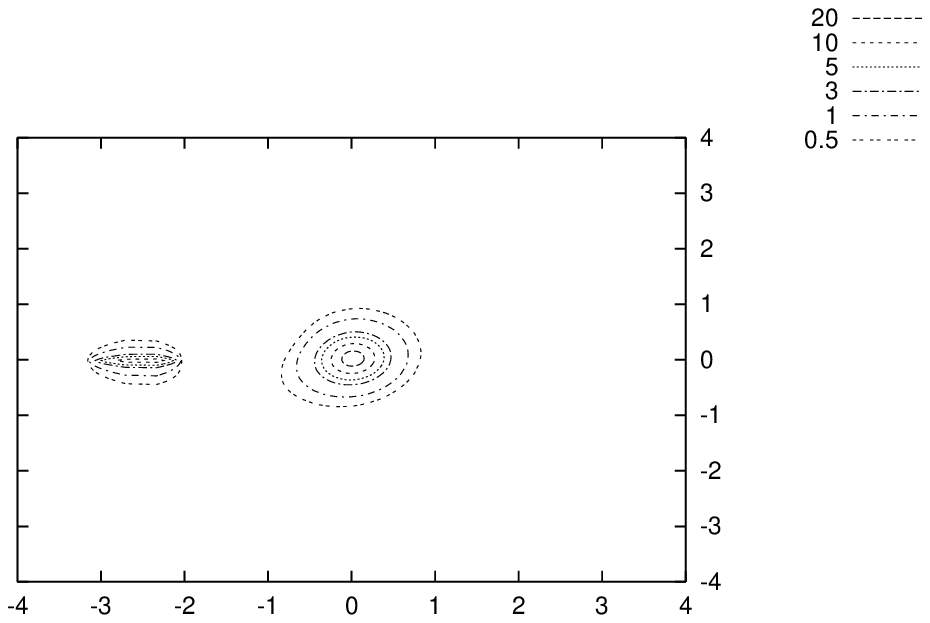}
\\
\includegraphics[scale=0.45]{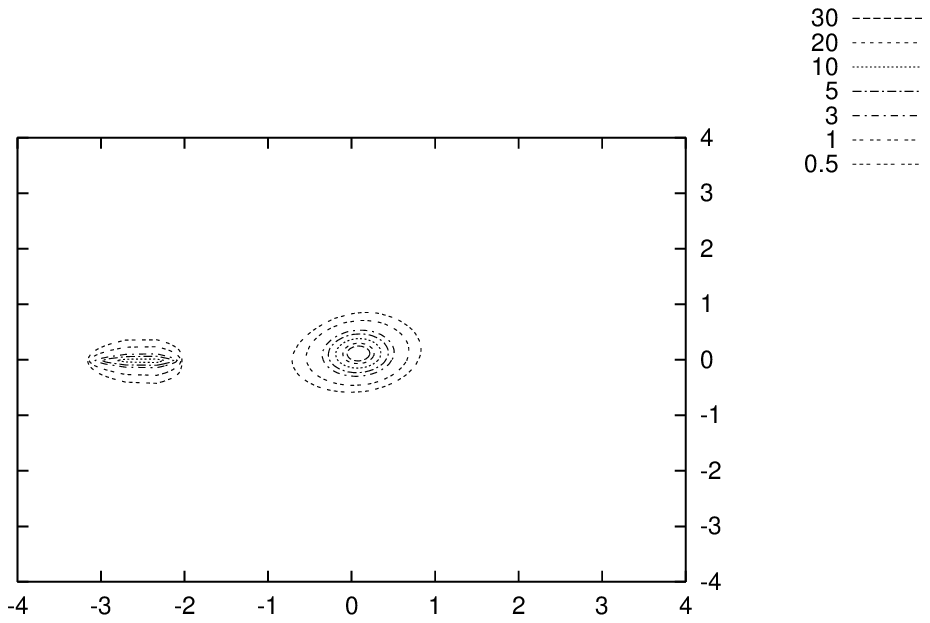}
\includegraphics[scale=0.45]{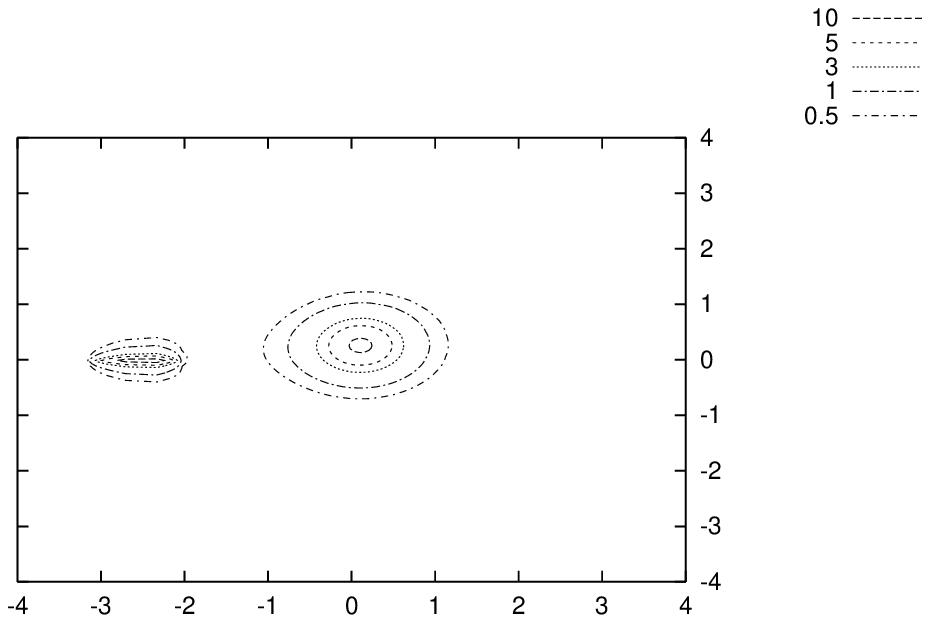}
\\
\includegraphics[scale=0.45]{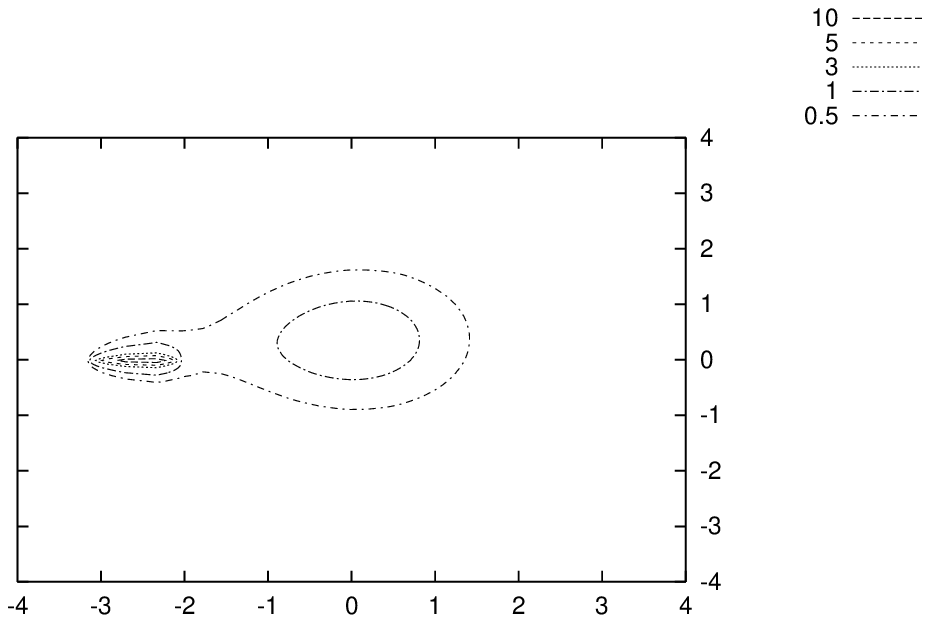}
\includegraphics[scale=0.45]{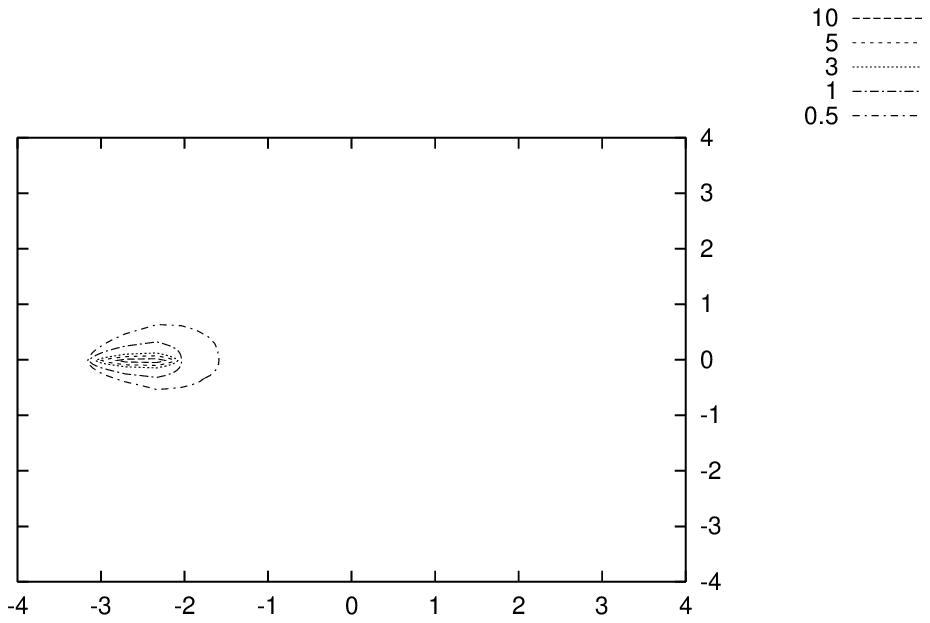}
\caption{Energy density for $\gamma = \frac{\pi}{4}$ at t=0, 2, 4, 6, 8 and 9.75}
\label{sspir}
\end{figure}

We also simulated the soliton defect system using the inhomogeneous ansatz (equation
(\ref{inhom})) as the initial condition, with the initial $\lambda = 2.5$, and the
initial soliton position relative to the defect at $\theta_{d}= 0$, $r_{d}= 2.5$, so
that the contribution to the ansatz from the soliton is significant compared to that
from the defect in the region of the soliton. With this ansatz the soliton spikes for
$\gamma = \frac{\pi}{2}$ (see figure (\ref{sspike})) and spreads at $\gamma =-
\frac{\pi}{2}$ (see figure (\ref{sspread})). If $\gamma = \frac{\pi}{4}$ in the initial
condition, the soliton starts to spike, but then later spreads (see figure(\ref{sspir})).
Table (\ref{senerg}) shows the energies for simulations with these initial conditions.
Again, these energies are only included for their value in comparing channels and to
show that energy is conserved during the free run of the simulation.
\begin{table}[h]
\begin{center}
\begin{tabular}{|c|c|c|c|c|}
\hline
$\gamma$     & Initial Energy & Energy after Relaxation & Time at end & Final \\
             &   ($t=0$)      &        ($t=0.5$)        & of Simulation & Energy \\
\hline
$\frac{\pi}{2}$ & 1.597 & 1.593 & 3.745 & 1.593 \\
\hline
0 & 1.523 & 1.516 & 9.995 & 1.516 \\
\hline
$-\frac{\pi}{2}$ & 1.442 & 1.437 & 9.995 & 1.438 \\
\hline
\end{tabular}
\caption{Energy of System in Inhomogeneous Ansatz}
\label{senerg}
\end{center}
\end{table}

\subsection{Collective Coordinate Approach}

To further the understanding of our numerical results we carried out some approximate analytic
work based on the so called collective coordinate approach. We substituted one of our ansatze into the
Lagrangian density, integrated over the spatial variables to find a true Lagrangian and then
considered one or more of the ansatz parameters as dynamic variables, so for example Lagrangian
(\ref{siglag}) together with
\begin{equation}
\vec\phi=\vec\phi(\lambda, \dot\lambda, x, y),
\end{equation}
implies that
\begin{equation}
\mathcal{L}= \mathcal{L}(\lambda, \dot\lambda, x, y),
\end{equation}
which may then be integrated over to give
\begin{equation}
L = \int \mathcal{L}(\lambda, \dot\lambda, x, y) dx dy = L(\lambda, \dot\lambda),
\end{equation}
allowing us to construct an equation of motion for $\lambda$.

This effectively constrains the solutions of the field theory to the submanifold defined by
the ansatz -- in the example above our solutions are constrained to a 2 dimensional submanifold
of the infinite dimensional phase space of $\vec\phi(x, y)$. This does not allow for soliton
unwinding or the release of radiation with our ansatze, as the ansatze never have radiation and
always have a soliton.

\clearpage

If we take the ansatz (\ref{wansa}) together with the Lagrangian (\ref{siglag}) expressed
in terms of stereographic coordinates
\begin{equation}
\mathcal{L}=\frac{\partial_{\mu}W\partial^{\mu}W^{*}}{4(1+|W|^{2})^{2}},
\end{equation}
we may construct the Lagrangian density in terms of position and a few parameters. If we
take the initial soliton and defect position outlined in section (\ref{steans}) then $W$
becomes
\begin{equation}
W = \frac{2\lambda(r\sin(\theta-\gamma)-2.5\sin\gamma)+i(1-\lambda^{2}(r^{2}+5r\cos\theta+6.25))}
{1+\lambda^{2}(r^{2}+5r\cos\theta+6.25)+2\lambda(r\cos(\theta-\gamma)+2.5\cos\gamma)}\tan\frac{\theta}{4},
\end{equation}
where ($r,\theta$) are polar coordinates, $\lambda$ parameterises the (inverse) width of the soliton and
$\gamma$ is the phase of the soliton. Note that the soliton has position ($2.5$, $\pi$) and the defect
is at the origin in this coordinate system.  Taking spatial derivatives of $W$ is
then a straight forward if somewhat tedious process. Time derivatives may be found by treating one or
more of the parameters as dynamic - if we consider $\lambda$ to be dynamic and the other parameters to
be static the $\partial_{t} = \dot\lambda\partial_{\lambda}$. This substitution of an ansatz and
use of parameters as dynamic variables is equivalent to assuming that the field configuration
moves quasi-statically from one configuration to another with different values of the dynamic variables.
This assumption is valid only if our ansatz is close to equilibrium and our dynamic parameter only
varies slowly with time (i.e. in this case $\dot\lambda$ is small).

Using this approximation our Lagrangian density becomes
\begin{equation}
\mathcal{L}=\frac{r^{2}\partial_{r}W\partial_{r}W^{*}+\partial_{\theta}W\partial_{\theta}W^{*}
-r^{2}\dot\lambda^{2}\partial_{\lambda}W\partial_{\lambda}W^{*}}{4r^{2}(1+|W|^{2})^{2}}.
\end{equation}
So our approximate Lagrangian with a time dependant $\lambda$ becomes
\begin{equation}
L=A(\lambda)-\dot\lambda^{2}B(\lambda),
\end{equation}
where
\begin{equation}
A(\lambda)=\int\frac{r^{2}\partial_{r}W\partial_{r}W^{*}+\partial_{\theta}W\partial_{\theta}W^{*}}{4r^{2}(1+|W|^{2})^{2}}rdrd\theta
\end{equation}
and
\begin{equation}
B(\lambda)=\int\frac{\partial_{\lambda}W\partial_{\lambda}W^{*}}{4(1+|W|^{2})^{2}}rdrd\theta.
\end{equation}
The Euler-Lagrange equation then gives us
\begin{equation}
\ddot\lambda=\frac{-(A'(\lambda)+\dot\lambda^{2}B'(\lambda))}{2B(\lambda)}.
\end{equation}

We considered the example above, where all variables are static except for $\lambda$,
but found an analytic integration to be intractable. We carried out the integration and time
evolution numerically, using a fourth order Runge-Kutta algorithm for the time evolution. We also
carried out a similar analysis using the inhomogeneous ansatz (\ref{inhom}).

\begin{figure}[hp]
\includegraphics[scale=0.45]{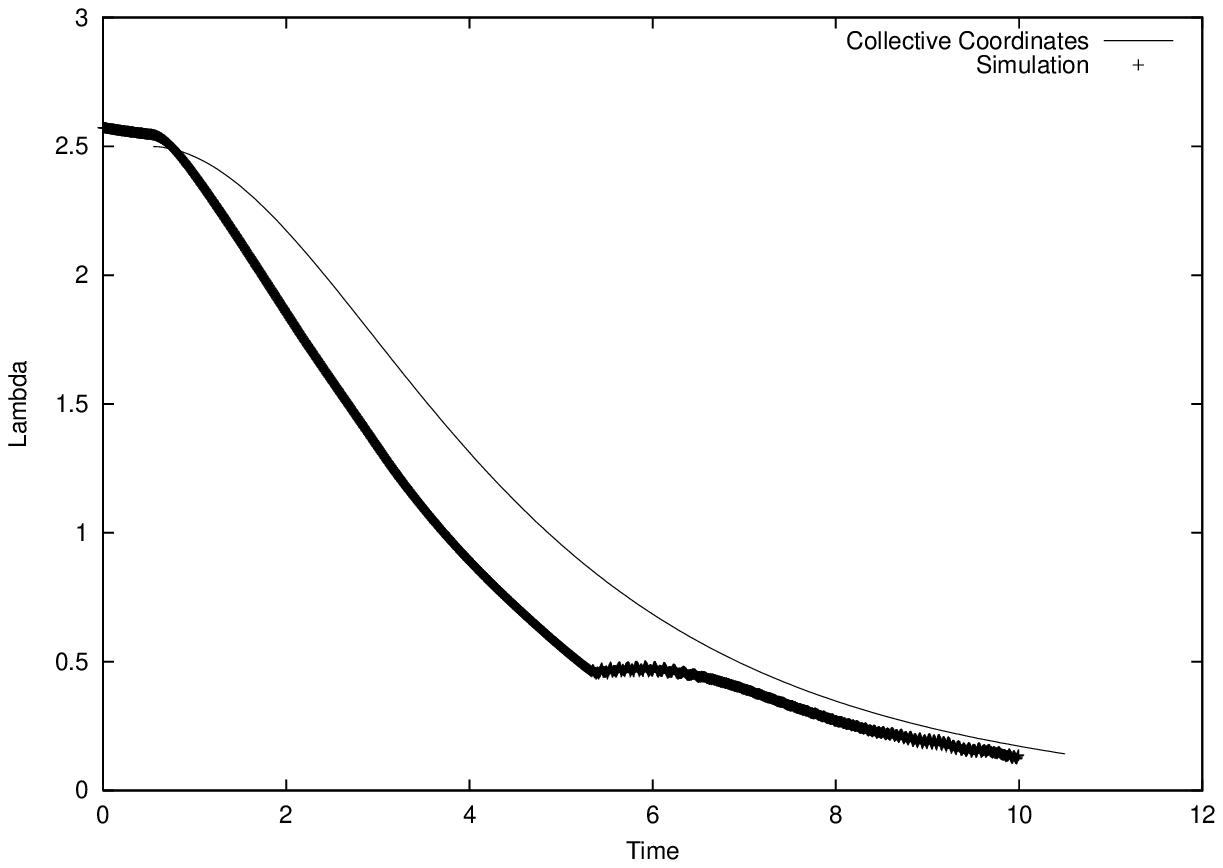}
\includegraphics[scale=0.45]{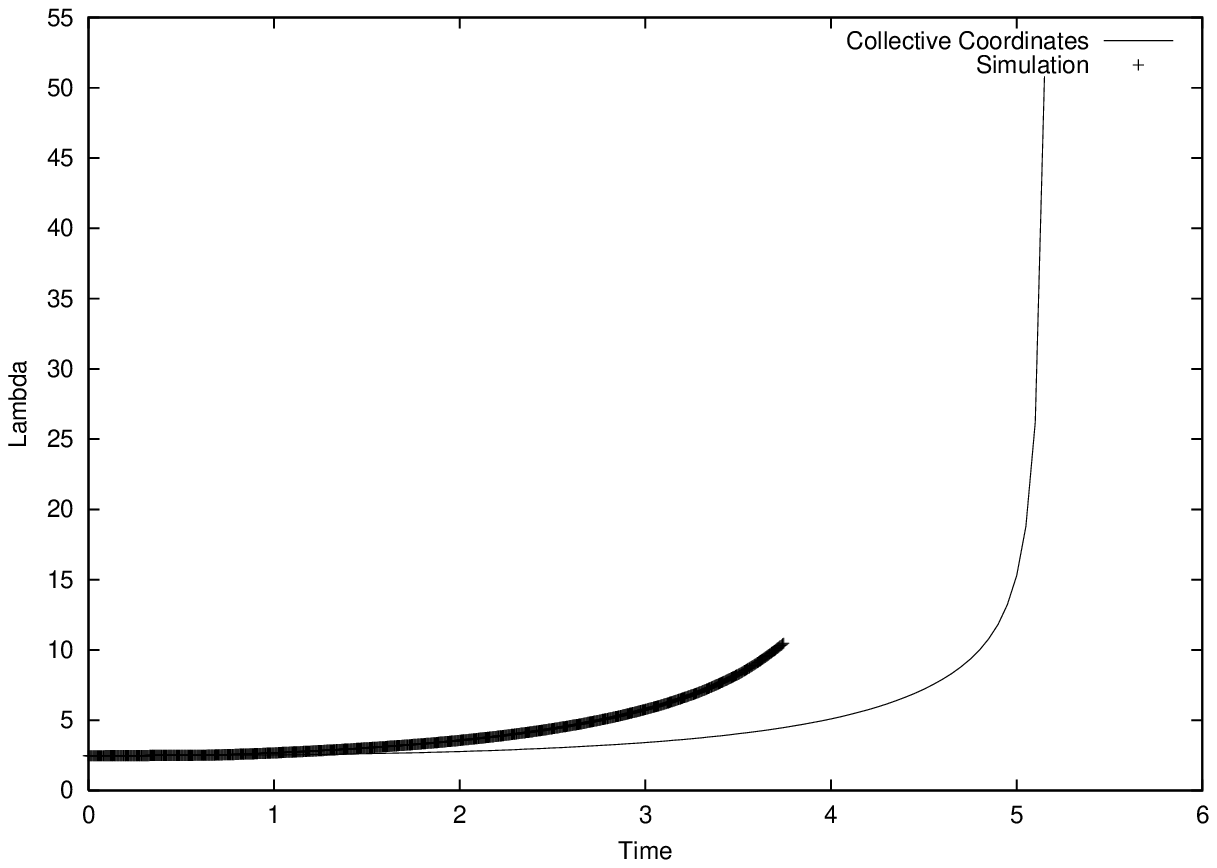}
\caption{$\lambda$ vs. time for spiking and broadening channels in the stereographic ansatz}
\label{geodesstere}
\end{figure}

\begin{figure}[hp]
\includegraphics[scale=0.45]{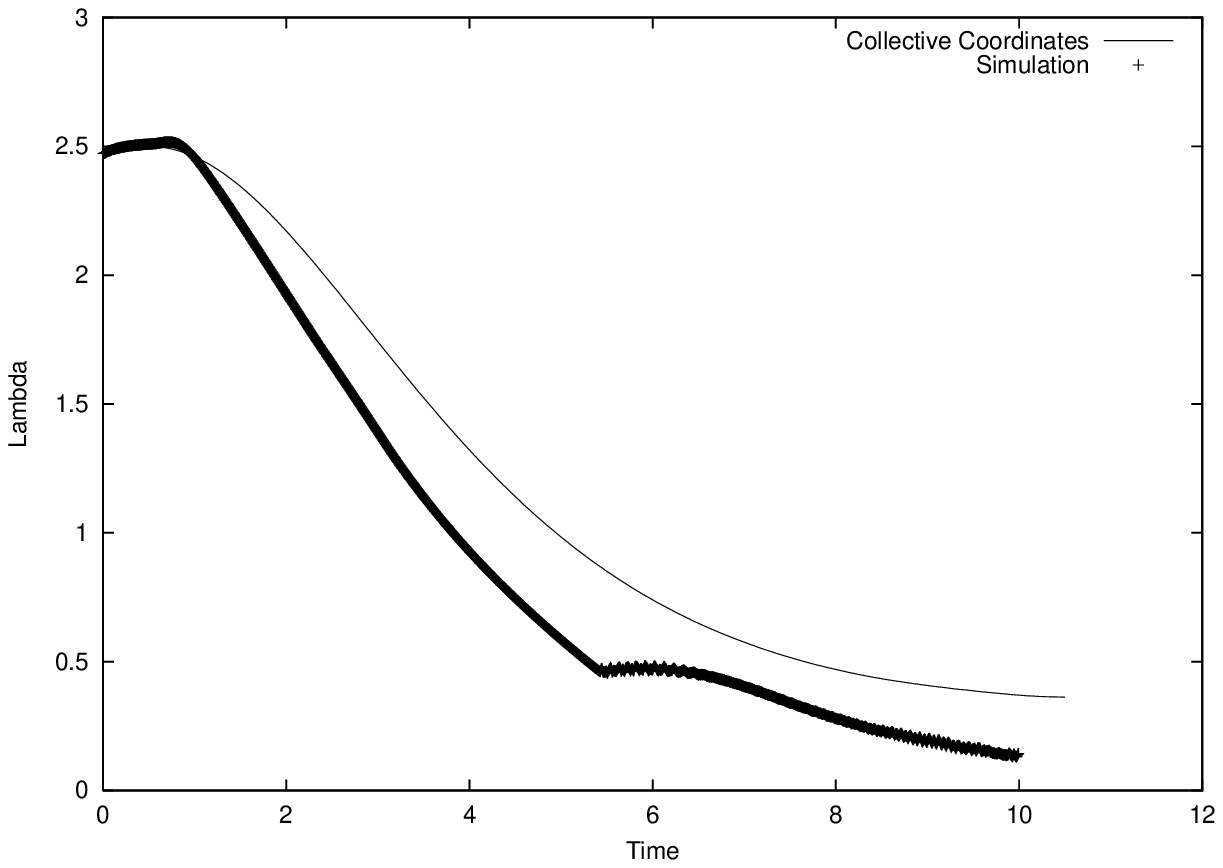}
\includegraphics[scale=0.45]{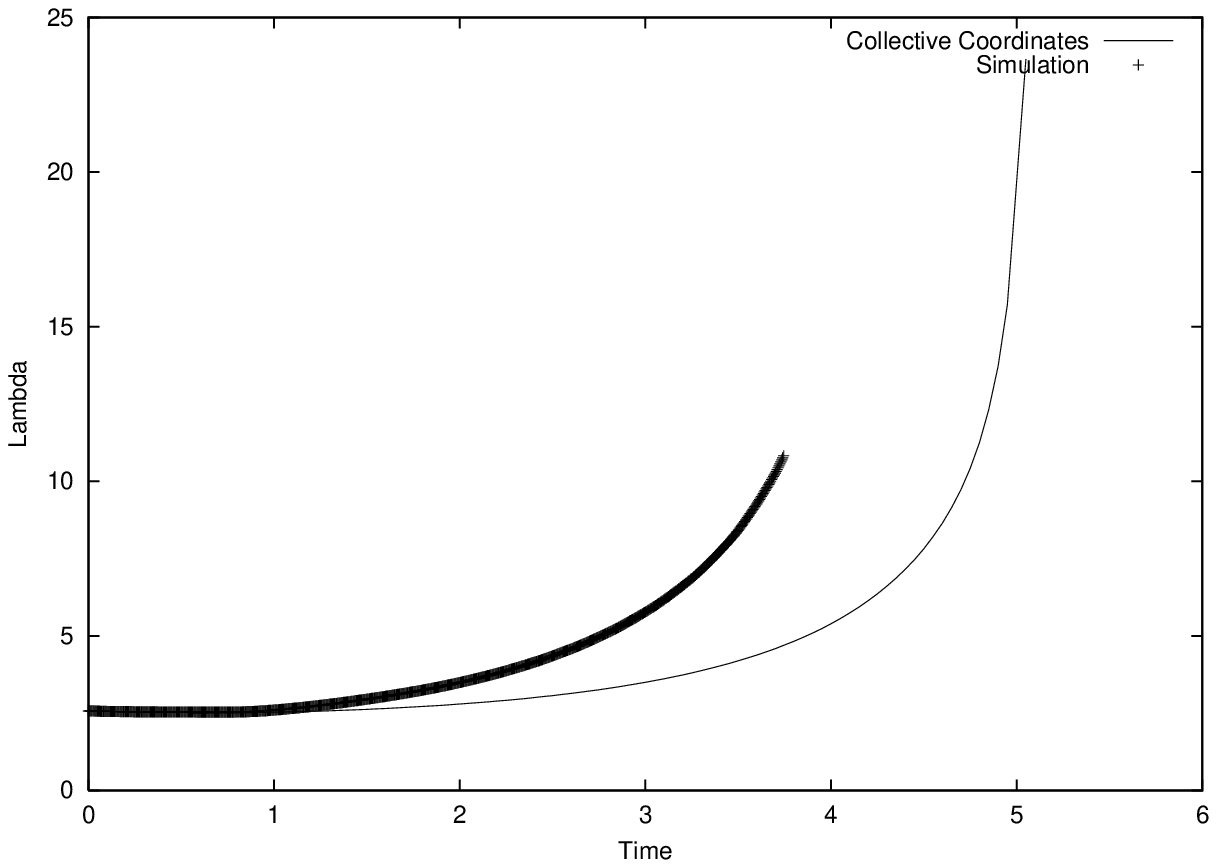}
\caption{$\lambda$ vs. time for spiking and broadening channels in the inhomogeneous ansatz}
\label{geodesinhom}
\end{figure}

In figures (\ref{geodesstere}, \ref{geodesinhom}) we show the results of this treatment against
those of the full simulation. The two treatments produce broadly similar results, although the
rapidity of the broadening and spiking is faster in the full simulation. This is not entirely
unreasonable considering that we have moved from around 160,000 degrees of freedom to 2! The
curves from the full simulation were found by finding the maximum winding number density on
the grid and then finding the value of lambda that would give a single soliton of
the form of equation (\ref{phisol}) this maximum
winding number density. One consequense of this technique is that for broad solitons the
maximum winding number for the soliton may be smaller than the maximum on the grid -- this
leads to the curves for the broadening channel becoming unreliable at around $t=5$. Naturally, the
collective coordinate simulation began at the point when the relaxation ended in the full
simulation.

\section{Conclusions}

We have examined the sigma model and baby Skyrme model with $\mathbb{R}P^{2}$ as a target
space and found these models to be identical to their $S^{2}$ counterparts in the absence
of defects. We examined the interaction between defects and soliton-like lumps in the
sigma model and found the interaction to depend on the relative phase of the lump. We
found a channel which causes the soliton to broaden and another which causes the soliton to
spike. When the soliton overlapped the defect significantly, the soliton would unwind. We
broadly reproduced this behaviour with the collective coordinate approach, using only two
collective coordinates.

A range of possibilities for future work present themselves -- studying the interaction
between two defects, the interactions of two lumps in the presence of a defect and the
behaviour of the $\mathbb{R}P^{2}$ sigma model on a torus all have the potential to exhibit
new and interesting behaviour.

\subsection*{Acknowledgements}

I would like to thank Wojtek Zakrzewski for his advice and guidance. I would also like to
thank PPARC for funding this research.

\end{document}